\documentclass[11pt,english,epsf]{scrartcl}
\usepackage{lmodern}
\usepackage[T1]{fontenc}
\usepackage[latin9]{inputenc}
\usepackage{geometry}
\usepackage{amsmath}
\usepackage{amssymb}
\usepackage{graphicx}

\newtheorem{theorem}{Theorem}

\newcommand {\dfn} {\stackrel{\Delta} {=}}
\newcommand {\exe} {\stackrel{\cdot} {=}}

\newcommand {\bsigma} {\mbox{\boldmath $\sigma$}}

\newcommand {\bu} {\mbox{\boldmath $u$}}
\newcommand {\bv} {\mbox{\boldmath $v$}}

\newcommand {\bx} {\mbox{\boldmath $x$}}
\newcommand {\by} {\mbox{\boldmath $y$}}

\newcommand {\bE} {\mbox{\boldmath $E$}}

\newcommand {\hP} {\hat{P}}
\newcommand {\hI} {\hat{I}}
\newcommand {\hH} {\hat{H}}

\newcommand {\tR} {\tilde{R}}

\newcommand {\bU} {\mbox{\boldmath $U$}}
\newcommand {\bV} {\mbox{\boldmath $V$}}

\newcommand {\bX} {\mbox{\boldmath $X$}}
\newcommand {\bY} {\mbox{\boldmath $Y$}}

\newcommand{\calA}{{\cal A}}

\newcommand{\calC}{{\cal C}}

\newcommand{\calE}{{\cal E}}

\newcommand{\calH}{{\cal H}}
\newcommand{\calI}{{\cal I}}

\newcommand{\calS}{{\cal S}}
\newcommand{\calT}{{\cal T}}
\newcommand{\calU}{{\cal U}}
\newcommand{\calV}{{\cal V}}

\newcommand{\calX}{{\cal X}}
\newcommand{\calY}{{\cal Y}}
\newcommand{\calZ}{{\cal Z}}
\allowdisplaybreaks

\begin{document}
\thispagestyle{empty}
\title{Universal Decoding for Source--Channel Coding with Side Information
\thanks{This research was partially supported by the Israel Science Foundation
(ISF), grant no.\ 412/12.}}
\author{Neri Merhav}
\date{}
\maketitle

\begin{center}
Department of Electrical Engineering \\
Technion - Israel Institute of Technology \\
Technion City, Haifa 32000, ISRAEL \\
E--mail: {\tt merhav@ee.technion.ac.il}\\
\end{center}
\vspace{1.5\baselineskip}
\setlength{\baselineskip}{1.5\baselineskip}

\begin{center}
{\bf Abstract}
\end{center}
\setlength{\baselineskip}{0.5\baselineskip}
We consider a setting of Slepian--Wolf coding, where the random bin of
the source vector undergoes channel coding,
and then decoded at the receiver,
based on additional side information, correlated to the source. For a given distribution of the
randomly selected channel codewords, we propose a universal decoder that
depends on the statistics of neither the correlated sources nor the
channel, assuming first that they are both memoryless. Exact analysis of the
random--binning/random--coding error exponent of this universal decoder shows
that it is the same as the one achieved by the optimal maximum a--posteriori
(MAP) decoder. Previously known results on universal Slepian--Wolf source decoding,
universal channel decoding, and universal source--channel decoding,
are all obtained as special cases of this result.
Subsequently, we further generalize the results in several directions,
including: (i) finite--state sources and finite--state channels, along with a
universal decoding metric that is based on Lempel--Ziv parsing, (ii) arbitrary sources
and channels, where the universal decoding is with respect to a given class of decoding
metrics, and (iii) full (symmetric) Slepian--Wolf coding, where both source streams are separately
fed into random--binning source encoders, 
followed by random channel encoders, which are then jointly decoded by
a universal decoder.

\setlength{\baselineskip}{1.5\baselineskip}
\newpage

\section{Introduction}

Universal decoding for unknown channels is a topic that attracted
considerable attention throughout the last four decades.
In \cite{Goppa75}, Goppa was the first to
offer the {\it maximum mutual
information} (MMI) decoder, which decodes the message as the one whose
codeword has the largest
empirical mutual information with the channel output sequence. Goppa
proved that for
discrete memoryless channels (DMC's), MMI
decoding attains capacity. Csisz\'ar and K\"orner \cite[Theorem 5.2]{CK81}
have further showed that
the random coding error exponent of the MMI decoder, pertaining to the
ensemble of the uniform
random coding distribution over a certain type class, achieves the same random
coding error exponent as the optimum, maximum likelihood (ML) decoder.
Ever since these early works on universal channel decoding, a considerably
large volume of research work has been done, see, e.g.,
\cite{Csiszar82}, \cite{FL98}, \cite{FM02}, \cite{LZ98}, \cite{LF12a},
\cite{LF12b}, \cite{Merhav93}, \cite{MW12}, 
for a non--exhaustive list of works on memoryless channels, as well as more general
classes of channels.

At the same time, considering the analogy between channel coding and
Slepian--Wolf (SW) source coding, it is not surprising that universal schemes
for SW decoding, like the {\it minimum entropy} (ME) decoder, have also been derived,
first, by Csisz\'ar and K\"orner \cite[Exercise 3.1.6]{CK81}, and later further developed by
others in various directions, see, e.g., \cite{CHJL08}, \cite{Draper04}, \cite{Kieffer80}, \cite{OH94}, 
\cite{SBB05}, \cite{WM15}.

Much less attention, however, has been devoted to universal decoding for joint
source--channel coding, where both the source and the channel are unknown to
the decoder. Csisz\'ar \cite{Csiszar80} was the first to propose such a universal decoder,
which he referred to as the {\it generalized MMI} decoder. The generalized MMI decoding
metric, to be maximized among all messages, is
essentially\footnote{Strictly speaking, Csisz\'ar's decoding 
metric is slightly different, but is asymptotically equivalent to
this definition.} given by the difference between the
empirical input--output mutual information of the channel
and the empirical entropy of the source. In a way, it naturally combines the
concepts of MMI channel decoding and ME source decoding.
But the emphasis in \cite{Csiszar80} was inclined much more towards upper and
lower bounds on the reliability function, whereas the universality of the
decoder was quite a secondary issue. Consequently, 
later articles that refer to \cite{Csiszar80} also focus, first and foremost,  on the
joint source--channel reliability function and not really on universal decoding.
We are not aware of subsequent works on universal source--channel decoding
other than \cite{MW06}, which concerns a
completely different setting, of zero--delay coding.

\begin{figure}[ht]
\vspace*{1cm}
\hspace*{2cm}\input{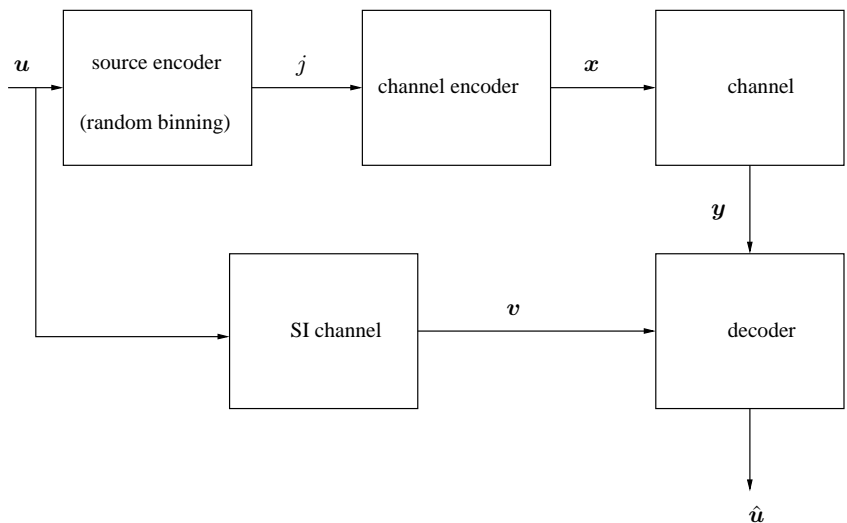}
\caption{\small Slepian--Wolf source coding, followed by channel coding. The
source $\bu$ is source--channel encoded, whereas the correlated SI $\bv$
(described as being generated by a DMC fed by $\bu$) is available at the
decoder.}
\label{sw1}
\end{figure}

In this work, we consider universal joint source--channel decoding in several
settings that are all more general than that of \cite{Csiszar80}. 
In particular, we begin by considering the
communication system depicted in Fig.\ \ref{sw1}, which is described as
follows: A source vector $\bu$, emerging from a discrete memoryless source
(DMS), undergoes Slepian--Wolf encoding (random binning) at rate $R$, followed by 
channel coding (random coding). The discrete memoryless channel (DMC) 
output $\by$ is fed into the decoder, along with a side
information (SI) vector $\bv$, correlated to the source $\bu$, and the output
of the decoder, $\hat{\bu}$, should agree with $\bu$ with probability as high as
possible. 

Our first step is to characterize the exact exponential rate of the expected
error probability, associated with the optimum MAP decoder, 
where the expectation is over both ensembles of the
random binning encoder and the random channel code. We refer to this
exponential rate as the {\it random--binning/random--coding error exponent}.
The second step, which is the more important one for our purposes,
is to show that this error exponent is also achieved by a
universal decoder, that depends neither on the
statistics of the source, nor of the channel, and which is similar to Csisz\'ar's generalized MMI
decoder. Beyond the fact this model is more general than the one in
\cite{Csiszar80} (in the sense of including the random binning component as
well as decoder SI), 
the assertion of the universal optimality of the
generalized MMI decoder is stronger here than in \cite{Csiszar80}. 
In \cite{Csiszar80} the performance of the generalized MMI decoder is compared
directly to an upper bound on the joint source--channel reliability function,
and the claim on the optimality of this decoder is asserted only in the range where
this bound is tight. Here, on the other hand (similarly as in earlier
works on universal pure channel decoding), we argue that the generalized
MMI decoder is {\it always} asymptotically as good as the optimal MAP decoder, in
the error exponent sense, no matter whether or not
there is a gap between the achievable exponent and the upper bound on the reliability function.
In other words, like in previous works on universal decoding, 
the focus is on asymptotic optimality of the decoder for the
average code and for an unknown channel, rather than on optimality of the overall communication
system. However, as we shall see later on, since full optimization of the random coding ensemble 
is infeasible, due to channel uncertainty, the best one can hope for is the
MAP source--channel error exponent due to Gallager \cite[Problem
5.16]{Gallager68}. We also provide an upper bound to the error exponent 
for any communication system with the configuration of Fig.\ \ref{sw1} and
discuss the conditions under which it is met.

One motivation for studying this model is that it captures, in a unified
framework, several
important special cases
of communication systems, from the perspective of universal decoding.
\begin{enumerate}
\item Separate source coding and channel coding without SI: letting $\bv$ be
degenerate.
\item Pure SW source coding: letting the channel be clean ($\by\equiv\bx$) 
and assuming the channel
alphabet to be very large (so that the probability for two or more identical
codewords would be negligible).
\item Pure channel coding: letting the source be binary and symmetric, and the
SI be degenerate.
\item Joint source--channel coding with and without SI: letting the binning rate
$R$ be sufficiently large, so that probability of ambiguous binning (i.e., when two
or more source vectors are mapped into the same bin) is negligible.
In this case, the mapping between source vectors and channel input vectors is
one--to--one with high probability, and therefore, this is a joint
source--channel code. More details on this aspect will follow in the sequel.
\item Systematic coding: letting the SI channel (from $\bu$ to $\bv$) in Fig.\ \ref{sw1} be identical
to the main channel (from $\bx$ to $\by$), and then the SI channel may represent transmission of the
systematic (uncoded) part of the code (see discussions on this point of view 
also in \cite{Merhav08} and \cite{SVZ98}).
\end{enumerate}

Another motivation is that it serves as the basis for the
more important part of the paper, where we provide 
three further extensions of this communication system model. In at least two of these
more general situations, the analysis is more tricky
and several difficulties that are encountered need to be handled with
care. The extended scenarios are the following.
\begin{enumerate}
\item Extending the scope from memoryless sources and channels to 
finite--state sources and finite--state channels. Here, the universal
joint source--channel decoding metric is based on Lempel--Ziv (LZ) parsing,
with the inspiration of \cite{Ziv85}. The non--trivial parts of the
analysis (not encountered in \cite{Ziv85} or other related works) 
are mainly those described in items 1, 7 and 8 of Subsection 5.1.
\item Further extending the scope to arbitrary sources and channels, but
allowing a given, limited
class of reference decoding metrics. We propose a universal joint source--channel
decoder with the property that,
no matter what the underlying source and channel
may be, this universal decoder is
asymptotically as good as the best decoder in the class for these source and
channel. This extends the recent study in \cite{givenclassofmetrics}, from pure
channel coding to joint source--channel coding.
\item Generalizing to the model to separate encodings (source binning followed
by channel coding) and joint decoding of two correlated sources (see Fig.\
\ref{sw2} in Section 5.3). Here the
universal decoder must handle several types of error events due to possible
ambiguities in the binning encoder. As a consequence of this fact, the
proposed universal decoding metric for this scenario is surprisingly different
from what one may expect.
\end{enumerate}

Finally, a few words are in order concerning the error exponent analysis. The
ensemble of codes in our setting combines random binning (for the source
coding part) and random coding (for the channel coding part), which is
considerably more involved than ordinary error exponent analyses that is
associated with either one but
not both. This requires a rather careful analysis, in two steps, where in the first, we take
the average probability of error over the ensemble of random binning codes, for
a given channel code, and at the second step, we average over the ensemble
of channel codes. The latter employs the type class enumeration
method \cite[Chap.\ 6]{fnt}, which has already proved 
rather useful as a tool for obtaining exponentially
tight random coding bounds in various contexts (see, e.g., 
\cite{REM1st}, \cite{swlist}, \cite{bid} for a sample),
and this work is no exception in that respect, as
the resulting error exponents are tight for the average code.

The remaining part of the paper is organized as follows.
In Section 2, we establish notation conventions, formalize the model and the
problem, and finally, review
some preliminaries. Section 3 provides the main result along with some
discussion. The proof of this result appears in Section 4, and finally,
Section 5 is devoted for the various extensions described above.

\section{Notation Conventions, Problem Setting and Preliminaries}

\subsection{Notation Conventions}

Throughout the paper, random variables will be denoted by capital
letters, specific values they may take will be denoted by the
corresponding lower case letters, and their alphabets
will be denoted by calligraphic letters. Random
vectors and their realizations will be denoted,
respectively, by capital letters and the corresponding lower case letters,
both in the bold face font. Their alphabets will be superscripted by their
dimensions. For example, the random vector $\bX=(X_1,\ldots,X_n)$, ($n$ --
positive integer) may take a specific vector value $\bx=(x_1,\ldots,x_n)$
in $\calX^n$, the $n$--th order Cartesian power of $\calX$, which is
the alphabet of each component of this vector.
Sources and channels will be denoted by the letter $P$, $Q$, or $W$,
subscripted by the names of the relevant random variables/vectors and their
conditionings, if applicable, following the standard notation conventions,
e.g., $Q_X$, $P_{Y|X}$, and so on. When there is no room for ambiguity, these
subscripts will be omitted. To avoid cumbersome notation, the various
probability distributions
will be denoted as above, no matter whether probabilities of single symbols or
$n$--vectors are addressed. Thus, for example, $P_U(u)$ (or $P(u)$) will denote the
probability of a single symbol $u\in\calU$, whereas  $P_U(\bu)$ (or $P(\bu)$)
will stand for the probability of the $n$--vector $\bu\in\calU^n$.
The probability of an event $\calE$ will be denoted by $\mbox{Pr}\{\calE\}$,
and the expectation
operator with respect to (w.r.t.) a probability distribution $P$ will be
denoted by $\bE\{\cdot\}$. 
The entropy of a generic distribution $Q$ on $\calX$ will be denoted by
$\calH(Q)$. For two
positive sequences $a_n$ and $b_n$, the notation $a_n\exe b_n$ will
stand for equality in the exponential scale, that is,
$\lim_{n\to\infty}\frac{1}{n}\log \frac{a_n}{b_n}=0$.
Accordingly, the notation $a_n\exe 2^{-n\infty}$ means that
$a_n$ decays at a super--exponential rate (e.g., double--exponentially).
Unless specified otherwise, logarithms and exponents, throughout this paper,
should be understood to be taken to the base 2.
The indicator function
of an event $\calE$ will be denoted by $\calI\{E\}$. The notation $[x]_+$
will stand for $\max\{0,x\}$. The minimum between two reals, $a$ and $b$, will
frequently be denoted by $a\wedge b$. The cardinality of a finite set, say
$\calA$, will be denoted by $|\calA|$.

The empirical distribution of a sequence $\bx\in\calX^n$, which will be
denoted by $\hat{P}_{\bx}$, is the vector of relative frequencies
$\hat{P}_{\bx}(x)$
of each symbol $x\in\calX$ in $\bx$.
The type class of $\bx\in\calX^n$, denoted $\calT(\bx)$, is the set of all
vectors $\bx'$
with $\hat{P}_{\bx'}=\hat{P}_{\bx}$. When we wish to emphasize the
dependence of the type class on the empirical distribution, say $Q$, we
will denote it by
$\calT(Q)$. Information measures associated with empirical distributions
will be denoted with `hats' and will be subscripted by the sequences from
which they are induced. For example, the entropy associated with
$\hat{P}_{\bx}$, which is the empirical entropy of $\bx$, will be
denoted\footnote{Note that here we use the letter $H$ in the ordinary font, as
opposed to the earlier defined notation of the entropy as a functional of a
distribution, where we used the calligraphic $\calH$.} by
$\hat{H}_{\bx}(X)$. Again, the subscript
will be omitted whenever it is clear from the context what sequence
the empirical distribution was extracted from.
Similar conventions will apply to the joint empirical
distribution, the joint type class, the conditional empirical distributions
and the conditional type classes associated with pairs of
sequences of length $n$.
Accordingly, $\hP_{\bx\by}$ would be the joint empirical
distribution of $(\bx,\by)=\{(x_i,y_i)\}_{i=1}^n$,
$\calT(\bx,\by)$ or $\calT(\hP_{\bx\by})$ will denote
the joint type class of $(\bx,\by)$, $\calT(\bx|\by)$ will stand for
the conditional type class of $\bx$ given
$\by$, $\hH_{\bx\by}(X,Y)$ will designate the empirical joint entropy of $\bx$
and $\by$,
$\hH_{\bx\by}(X|Y)$ will be the empirical conditional entropy,
$\hI_{\bx\by}(X;Y)$ will
denote empirical mutual information, and so on.

\subsection{Problem Setting for the Basic Setting}

Let $(\bU,\bV)=\{(U_t,V_t)\}_{t=1}^n$ be $n$ independent copies of a
pair of random variables, $(U,V)\sim
P_{UV}$, taking on values in finite alphabets, $\calU$ and $\calV$, respectively.
The vector $\bU$ will designate the source vector to be encoded, whereas the
vector $\bV$ will serve as correlated SI, available to the decoder.
Let $W$ designate a DMC, with single--letter, input--output transition
probabilities $W(y|x)$, $x\in\calX$, $y\in\calY$, $\calX$ and $\calY$ being
finite input and output alphabets, respectively. When the channel is fed by an
input vector $\bx\in\calX^n$, it produces\footnote{Without essential loss of
generality, and similarly as in \cite{Csiszar80}, 
we assume that the source and the channel operate at the
same rate, so that while the source emits the $n$--vector $(\bU,\bV)$,
the channel is used $n$ times exactly, transforming $\bx\in\calX^n$ to
$\by\in\calY^n$. The extension to the case where operation rates are different
(bandwidth expansion factor different from 1) is straightforward but is
avoided here, in the quest of keeping notation and expressions less
cumbersome.}
a channel output vector
$\by\in\calY^n$, according to 
\begin{equation}
W(\by|\bx)=\prod_{t=1}^n W(y_t|x_t).
\end{equation}
Consider the communication system depicted in Fig.\ \ref{sw1}.
When a given realization $\bu=(u_1,\ldots,u_n)$, of the source vector 
$\bU$, is fed into the system, it is encoded into one out of $M=2^{nR}$
bins, selected independently at random for every member of $\calU^n$. 
Here, $R>0$ is referred to as the {\it binning rate}. The bin
index $j=f(\bu)$ is mapped into a channel input vector
$\bx(j)\in\calX^n$, which in turn is transmitted across the channel $W$.
The various codewords $\{\bx(j)\}_{j=1}^M$ are selected independently at random
under the uniform distribution across a given type class $\calT(Q)$,
$Q$ being a given probability distribution over $\calX$.\footnote{Rather than
the same type $\calT(Q)$ for all bins, a more
general ensemble may allow different types of codewords to bins of different types of
source vectors. We will address this point in the next section.}
The randomly chosen codebook
$\{\bx(1),\bx(2),\ldots,\bx(M)\}$ will be denoted by $\calC$.
Both the channel encoder, $\calC$, and the realization of the random binning 
source encoder, $f$, are revealed to the decoder as well.
With a slight abuse of notation, we will sometimes denote $\bx(j)=\bx[f(\bu)]$
by $\bx[\bu]$. The optimal (MAP) decoder estimates $\bu$, using the channel
output $\by=(y_1,\ldots,y_n)$ and the SI vector $\bv=(v_1,\ldots,v_n)$,
according to:
\begin{equation}
\label{mapdecoder}
\hat{\bu}=
\mbox{arg}\max_{\bu} P(\bu,\bv)W(\by|\bx[\bu]).
\end{equation}
The average probability of error, $\overline{P_{\mbox{\tiny e}}}$, 
is the probability of the
event $\{\hat{\bU}\neq \bU\}$, where in addition to the randomness of
$(\bU,\bV)$ and the channel output $\bY$, the randomness of the source binning
code and the channel code are also taken into account. The
random--binning/random--coding error exponent, associated with the optimal, MAP decoder,
is defined as 
\begin{equation}
E(R,Q)=\lim_{n\to\infty}\left[-\frac{\log \overline{P_{\mbox{\tiny
e}}}}{n}\right],
\end{equation}
provided that the limit exists (a fact that will become evident from the analysis
in the sequel). 

The first step is to derive a single--letter expression for the exact 
random--binning/random--coding error exponent $E(R,Q)$.
While the MAP decoder depends on the source $P$ and the
channel $W$, the second step is to propose a universal decoder,
independent of $P$ and $W$, whose average error probability decays
exponentially at the same rate, $E(R,Q)$. Note that we are considering a fixed
$Q$ without attempt to maximize $E(R,Q)$ w.r.t.\ $Q$ since the maximizing $Q$
normally depends on the unknown channel $W$ (more on this in the next
subsection). Finally, our main goal is to extend the scope
beyond memoryless systems, as well as 
to the setting where the role of $\bV$ is no longer merely to serve as SI at the decoder, but
rather as another source vector, encoded similarly, but separately from $\bU$ (see Fig.\ \ref{sw2}).

\subsection{Preliminaries -- the Joint Source--Channel Error Exponent}

To the best of our knowledge, the first to consider error exponents for joint
source--channel coding (without SI) was Gallager (see also Jelinek
\cite{Jelinek68}). In the
second part of Problem 5.16 in his
textbook \cite{Gallager68} (pp.\ 534--535), the reader is requested to prove that
for a given DMS $P$, a given DMC $W$, and a given product distribution $Q$
for random selection of a channel input vector $\bx[\bu]$ for each source
vector $\bu$, the average probability of error is upper bounded by
\begin{equation}
\overline{P}_{\mbox{\tiny
e}}\le\exp\left\{-n\max_{0\le\rho\le 1}\left[E_0(\rho,Q)-
(1+\rho)\ln\left(\sum_{u\in\calU}[P(u)]^{1/(1+\rho)}\right)\right]\right\},
\end{equation}
where $E_0(\rho,Q)$ is the well--known Gallager function
\begin{equation}
E_0(\rho,Q)=-\ln\sum_{y\in\calY}\left[\sum_{x\in\calX}Q(x)W(y|x)^{1/(1+\rho)}\right]^{1+\rho}.
\end{equation}
It is easy to show (see Appendix) that this exponential upper bound is equivalent
to
\begin{equation}
\label{csiszargallager}
\overline{P}_{\mbox{\tiny
e}}\le\exp\left\{-n\min_{\calH(P)\le \tR\le\log|\calU|}\left[E^{\mbox{\tiny
s}}(\tR)+E_{\mbox{\tiny r}}^{\mbox{\tiny
c}}(\tR,Q)\right]\right\},
\end{equation}
where $E^{\mbox{\tiny s}}(\tR)$ is the source reliability function
\cite{Marton74}, given by
\begin{equation}
\label{srcexp}
E^{\mbox{\tiny s}}(\tR)=\min_{\{P':~\calH(P')\ge \tR\}} D(P'\|P),
\end{equation}
$D(P'\|P)$ being the Kullback--Leibler 
divergence between $P'$ and $P$, and
\begin{equation}
E_{\mbox{\tiny r}}^{\mbox{\tiny c}}(\tR,Q)=
\max_{0\le\rho\le 1}[E_0(\rho,Q)-\rho \tR].
\end{equation}

Slightly more than a decade later, Csisz\'ar
\cite{Csiszar80} derived upper and lower bounds on the
reliability function of lossless joint source--channel coding (again, without SI). 
Csisz\'ar has shown,
in that paper, that the reliability function, $E_{\mbox{\tiny jsc}}$, of lossless joint
source--channel coding is upper bounded by
\begin{equation}
\label{jscu}
E^{\mbox{\tiny jsc}} \le\min_{\calH(P)\le \tR\le\log|\calU|}
[E^{\mbox{\tiny s}}(\tR)+E^{\mbox{\tiny c}}(\tR)]
\end{equation}
where $E^{\mbox{\tiny c}}(\tR)$ is the channel reliability function, for which
there is a closed--form expression available only at rate zero (the zero--rate
expurgated exponent) and at rates
above the critical rate (the sphere--packing exponent).
The lower bound in \cite{Csiszar80} is given by
\begin{equation}
\label{jscl1}
E^{\mbox{\tiny jsc}} \ge\min_{\tR}[E^{\mbox{\tiny s}}(\tR)+E_{\mbox{\tiny
r}}^{\mbox{\tiny c}}(\tR)],
\end{equation}
where $E_{\mbox{\tiny r}}^{\mbox{\tiny c}}(\tR)$ is the random coding error
exponent of the channel $W$ and where we have relaxed the constraint on the
range of $\tR$ since this unconstrained minimum is attained in that range
anyway (see \cite[p.\ 323, one line before the Remark]{Csiszar80}).
The upper and the lower bounds coincide
(and hence provide the exact reliability function)
whenever the minimizing $\tR$, of the upper bound (\ref{jscu}),
exceeds the critical rate of the channel $W$. 

Note that the difference between
the achievable exponents of Gallager and Csisz\'ar is in the channel error
exponent terms. In the former, it is $E_{\mbox{\tiny r}}^{\mbox{\tiny c}}(\tR,Q)$, whereas in the
latter it is improved to 
$E_{\mbox{\tiny r}}^{\mbox{\tiny c}}(\tR)=\max_QE^{\mbox{\tiny c}}(\tR,Q)$. The reason
is that, while Gallager uses the same type random coding distribution $Q$ for
all codewords $\{\bx[\bu]\}$, Csisz\'ar partitions the source space
according to the
various types $\{P'\}$ and maps each such type into a channel subcode whose rate
is essentially $\tR=\calH(P')$ and for which the channel input type $Q$ is optimized according to
$\max_QE_{\mbox{\tiny r}}^{\mbox{\tiny c}}(\calH(P'),Q)$. The difference disappears, of course,
if for the channel $W$, the same $Q$ maximizes $E^{\mbox{\tiny c}}(\tR,Q)$ for
every $\tilde{R}$. This happens, for example, for
the modulo--additive channel $Y=X\oplus N$ ($N$ being independent of $X$),
where the uniform distribution $Q$ is optimal independently of $\tR$.

An expression equivalent to (\ref{jscl1}) is given by
\begin{equation}
\label{jscl2}
E^{\mbox{\tiny jsc}} \ge
\min_{P'}\max_Q\min_{W'}\{D(P'\|P)+D(W'\|W|Q)+[I(X;Y')-H(U')]_+\}
\end{equation}
where $U'$ is an auxiliary random variable drawn by a source $P'$ over $\calU$
(hence $H(U')=\calH(P')$), $X$ is governed by
$Q$, $Y'$ designates the output of an auxiliary channel $W':\calX\to\calY$ fed
by $X$, and $D(W'\|W|Q)$ is the Kullback--Leibler divergence between $W'$ and
$W$, weighted by $Q$, that is
\begin{equation}
D(W'\|W|Q)=\sum_{x\in\calX}Q(x)\sum_{y\in\calY}W'(y|x)\log\frac{W'(y|x)}{W(y|x)}.
\end{equation}
Here, the term $D(P'\|P)$ is parallel to the source coding exponent, $E^{\mbox{\tiny
s}}(\tR)$, whereas the sum of other two terms
can be referred to the channel coding exponent $E_{\mbox{\tiny
r}}^{\mbox{\tiny c}}(\tR)$ (see \cite{Csiszar80}).
This is true since the minimization over $P'$ in (\ref{jscl2}) can be carried
out in two steps, where in the first, one minimizes over all $\{P'\}$ with a given
entropy $\calH(P')=\tR$ (thus giving rise to $E^{\mbox{\tiny s}}(\tR)$ according to
(\ref{srcexp})), and then minimizes over $\tR$. 

In a nutshell, the idea behind the converse part in \cite{Csiszar80} is that
each type class $P'$, of source vectors, can be thought of as being mapped
by the encoder into a separate channel subcode at rate $\tR=\calH(P')$, and then the probability of
error is lower bounded by the contribution of the worst subcode.
This is to say that for the purpose of the lower bound, 
only decoding errors {\it within} each subcode are counted, whereas
errors caused by confusing two
source vectors that belong two different subcodes, are ignored. 
An interesting point, in this context, is that whenever the upper and the lower bound coincide
(in the exponential scale), this means that confusions {\it within} the subcodes
dominate the error probability, at least as far as error exponents are
concerned, whereas errors of confusing codewords from different subcodes are
essentially immaterial. We will witness the same phenomenon from a different
perspective, in the sequel.

For the achievability part of \cite{Csiszar80}, Csisz\'ar analyzes the
performance of a universal decoder, that is asymptotically equivalent to
the following:
\begin{equation}
\label{gmmicsiszar}
\hat{\bu}=\mbox{arg}\max_{\bu}[\hI_{\bx[\bu]\by}(X;Y)-\hH_{\bu}(U)].
\end{equation}
As mentioned earlier, Csisz\'ar refers to his decoder as the generalized MMI decoder.
An important point to observe, however, is that in this universal setting, it
makes sense to assume that the encoder does not know the channel $W$ either and
hence cannot match the optimal channel input type $Q$ to every given source
type rate $\tR=\calH(P')$ as described above, because this optimal type depends
on the unknown channel $W$ (see \cite[page 323, Remark]{Csiszar80}). Csisz\'ar
suggests, in this case, to select a fixed type $Q$ for all source types, in
which case the achievable exponent becomes the same as Gallager's exponent. Throughout
this paper, we shall adopt this suggestion, and for this reason, we have defined
our objective (in the previous section) to universally achieve $E(R,Q)$ for a given $Q$ without
attempt to optimize $Q$, 

\section{Results for the System of Fig.\ \ref{sw1}}

Our problem setting and results are more general than those of
\cite{Csiszar80} from the following aspects:
(i) we include side information $\bV$, (ii) we include a cascade of random binning
encoder and a channel encoder (separate source-- and channel coding), 
and (iii) we compare the performance of the universal decoder to
that of the MAP decoder (\ref{mapdecoder}) and show that they {\it always}
(i.e., even when the random coding ensemble is not good enough to achieve the
reliability function)
have the same error exponent, whereas Csisz\'ar compares
the performance of (\ref{gmmicsiszar}) to the upper bound (\ref{jscu}) and thus
may conclude for asymptotic optimality 
of the decoder (together the encoder) only when the exact joint
source--channel reliability function
is known.

Concerning (ii), one may wonder what is the motivation for
separate source-- and channel coding, because joint source--channel coding is always at
least as good. The answer to this question is two--fold:
\begin{enumerate}
\item In some applications, system constraints dictate
separate source-- and channel coding, for example, when the two encodings are
performed at different units/locations or 
when general engineering considerations (like modularity) dictate
separation. 
\item The joint source--channel setting, for a fixed channel input type $Q$,
can always be obtained as a
special case, by choosing the binning rate $R$ sufficiently high, since then the
binning encoder is a one--to--one mapping with an overwhelmingly high
probability and the channel code in cascade to the binning code is equivalent
to a direct mapping between source vectors and channel input vectors.
\end{enumerate}

Our main result is given by the following theorem.
\begin{theorem}
Consider the problem setting defined in Subsection 2.2.
\begin{itemize}
\item[(a)] The random--binning/random--coding error
exponent of the MAP decoder is given by
\begin{equation}
E(R,Q)=\min_{P_{U'V'},W'}\{D(P_{U'V'}\|P_{UV})+D(W'\|W|Q)+
[R\wedge I(X;Y')-H(U'|V')]_+\}
\end{equation}
where $(U',V')\in\calU\times\calV$ 
are auxiliary random variables jointly distributed according
to $P_{U'V'}$, 
and $Y'\in\calY$ is another auxiliary random variable that designates the
output of channel $W'$ when fed by $X\sim Q$.
\item[(b)] The universal decoders,
\begin{equation}
\label{gmmi}
\hat{\bu}=\mbox{arg}\max_{\bu}[\hI_{\bx[\bu]\by}(X;Y)-\hH_{\bu\bv}(U|V)]
\end{equation}
and
\begin{equation}
\label{gmmi2}
\hat{\bu}=\mbox{arg}\max_{\bu}[R\wedge
\hI_{\bx[\bu]\by}(X;Y)-\hH_{\bu\bv}(U|V)],
\end{equation}
both achieve $E(R,Q)$.
\end{itemize}
\end{theorem}

Decoder (\ref{gmmi}) is, of course, a natural
extension of (\ref{gmmicsiszar}) to our setting. As for (\ref{gmmi2}), while 
it offers no apparent advantage over
(\ref{gmmi}), it is given here as an alternative
decoder for future reference. It will turn out later that conceptually, (\ref{gmmi2}) lends
itself more naturally to the extension that deals with separate encodings
and joint decoding of two correlated sources, where in the extended
version of (\ref{gmmi2}), it will not be obvious (at least not to the author) 
that the operator  $R\wedge
(\cdot)$ is neutral (i.e., an expression like 
$R\wedge x$ can be simply replaced by $x$, as is indeed suggested here by
the equivalence between (\ref{gmmi}) and (\ref{gmmi2})).
Another interesting point concerning (\ref{gmmi2}),
is that it appears more clearly as a joint extension of the MMI decoder of
pure channel decoding and the ME decoder of pure source coding. When $R$
dominates the term $R\wedge \hI_{\bx[\bu]\by}(X;Y)$, the source coding
component of the problem is more prominent and (\ref{gmmi2}) is essentially
equivalent to the ME decoder. Otherwise, it is essentially equivalent to (\ref{gmmi}).

As can be seen, $E(R,Q)$ is monotonically non--decreasing in $R$,\footnote{
This fact is not completely trivial, since an increase in $R$ improves the
source binning part, but one may expect that it harms the channel coding part.
Nonetheless, as will become apparent 
in the sequel (see footnote 5), the combined effect of source binning
and channel coding gives a non--decreasing exponent as a function of $R$.}
but when $R$ is sufficiently large, 
the term $R\wedge I(X;Y')$ is dominated by $I(X;Y')$ (say, $R=\log|\calX|$), 
which yields saturation
of $E(R,Q)$ to the level of the joint
source--channel random coding exponent (for a given $Q$), 
similarly as in (\ref{jscl2}), except that here, the entropy $H(U')$ is
replaced by the conditional entropy $H(U'|V')$, due to the SI.
Obviously, if $V'$ is degenerate (e.g., equal to a fixed $v\in\calV$ with
probability one), then we are back to (\ref{jscl2}).
For another extreme case, 
if the channel is clean, $Q$ is uniform, and the channel alphabet
is very large, 
then $\hat{I}(X;Y')=\hH(X)=\log|\calX|$ is large as well, and
then $R\wedge \hI(X';Y)$ is dominated by $R$. In this case, we
recover the SW random binning error exponent (see, e.g., \cite{WM15} and
references therein).

Finally, although not directly related to the aspect of universality,
for the sake of completeness (and in analogy to
\cite{Csiszar80}), we next provide also a
converse bound that applies to any communication system of the
type depicted in Fig.\ \ref{sw1}, where both the binning code
$f:\calU^n\to\{1,2,\ldots,M\}$ and the channel code, that maps
$\{1,2,\ldots,M\}$ into $\calC$, are arbitrary (and
deterministic), and where the optimal MAP decoder is used.
It is not difficult to extend Lemma 2 of \cite{Csiszar80} to the scenario
under discussion and to argue that given the source $P_{UV}$, the channel $W$,
and the binning rate $R$, the highest achievable source--channel error
exponent, $E(P_{UV},W,R)$ is upper bounded by
\begin{equation}
E(P_{UV},W,R)\le
\min_{P_{U'V'},W'}\left\{D(P_{U'V'}\|P_{UV})+\calI\{H(U'|V')\le R\}\cdot
E^{\mbox{\tiny c}}[H(U'|V')]\right\},
\end{equation}
which follows immediately from the simple consideration of viewing each
conditional type $\calT(\bu'|\bv')$ (whose weight is exponentially
$2^{-nD(P_{U'V'}\|P_{UV})}$) as being encoded by a channel subcode at rate $\tR=H(U'|V')$, which is
the corresponding empirical conditional entropy. Now, as long as $\tR\le R$,
this conditional type may be mapped into the channel code without loss of
information and then the error probability within this subcode 
is lower bounded by $2^{-n[E^{\mbox{\tiny
c}}(\tR)+o(n)]}$ (as all source messages originating from $\calT(\bu'|\bv')$ are equally
likely given $\bv'$).
However, if $\tR > R$ most of the members of $\calT(\bu'|\bv)$ 
are mapped ambiguously by the binning encoder and the probability of error
goes to unity even without the channel noise, hence the factor
$\calI\{H(U'|V')\le R\}$ in the second term. If we further upper bound $E^{\mbox{\tiny
c}}(\cdot)$ by the sphere--packing exponent, then 
for the second term of the above we have
\begin{equation}
\calI\{H(U'|V')\le R\}\cdot
E_{\mbox{\tiny sp}}^{\mbox{\tiny c}}[H(U'|V')]=\max_Q\min\{D(W'\|W|Q):~R\wedge I(X;Y')\le H(U'|V')\}.
\end{equation}
To see why this is true, one simply examines the two cases, $H(U'|V')\le R$
and $H(U'|V') > R$. In the former case, the constraint $R\wedge I(X;Y')\le
H(U'|V')$ is equivalent to $I(X;Y')\le H(U'|V')$, and then both the
right--hand side (r.h.s.) and the left--hand side (l.h.s.) become 
$E_{\mbox{\tiny sp}}^{\mbox{\tiny c}}[H(U'|V')]$. In the
latter case, the constraint is trivially met for every $W'$, including
the choice $W'=W$ for which the r.h.s.\ vanishes, exactly like the l.h.s.
Putting this together, we have
\begin{equation}
E(P_{UV},W,R)\le
\min_{P_{U'V'}}\max_Q\min_{\{W':~R\wedge I(X;Y')\le H(U'|V')\}}
[D(P_{U'V'}\|P_{UV})+D(W'\|W|Q)].
\end{equation}
$E(R,Q)$ of Theorem 1 meets this upper bound whenever the minimizing
$P_{U'V'}$ and $W'$ of $E(R,Q)$ are such that $R\wedge I(X;Y')\le H(U'|V')$
and that the minimization over $P_{U'V'}$ can be interchanged with
the maximization over $Q$, e.g., when the optimal $Q$ is independent
of the coding rate, as discussed above.\footnote{Note that here, as opposed
to \cite{Csiszar80}, the independence of the optimal $Q$ upon $\tR$ 
is needed (in order to match the converse)
even when the channel is {\it known}, because the encoder is unaware
of the virtual rate $\tR=H(U'|V')$ due to the 
unavailability of $\bv$ at the encoder.}
The first condition can be 
stated differently as follows: if we formally define
$\tilde{E}^{\mbox{\tiny s}}(\tR)=
\min\{D(P_{U'V'}\|P_{UV}):~H(U'|V')=\tR\}$ (which
depends solely on the source) and
$\tilde{E}^{\mbox{\tiny c}}(\tR,R,Q)=
\min_{W'}\{D(W'\|W|Q)+[R\wedge
I(X;Y')-\tR]_+\}$ (which depends solely on the channel),
then the upper bound is attained if the value of $\tR$  
that minimizes $[E^{\mbox{\tiny
s}}(\tR)+E_{\mbox{\tiny r}}^{\mbox{\tiny c}}(\tR,R,Q)]$ is large enough such
that $E_{\mbox{\tiny r}}^{\mbox{\tiny c}}(\tR,R,Q)$ is achieved by $W'$
for which $R\wedge I(X';Y)\le \tR$.

\section{Proof of Theorem 1}

The outline of the proof is as follows.
We begin by showing that $E(R,Q)$ is an
upper bound on the error exponent associated with the MAP decoder, and then we
show that both universal decoders (\ref{gmmi}) and (\ref{gmmi2}) attain
$E(R,Q)$. The combination of these two facts will prove both parts of Theorem
1 at the same time.

As a first step, let the channel codebook $\calC$, as
well as the vectors
$\bu$, $\bv$, $\bx=\bx[\bu]$ and $\by$ be given, and let
$\overline{P}_e(\bu,\bv,\bx,\by,\calC)$ be the average error probability given
$(\bu,\bv,\bx,\by,\calC)$, where the averaging is w.r.t.\ the ensemble of
random binning source codes. For a given $\bu'\ne \bu$, let us define the set
\begin{equation}
\calA(\bu,\bu',\bv,\bx,\by)=\calT(Q)\bigcap
\left\{\bx':~P(\bu',\bv)W(\by|\bx')\ge P(\bu,\bv)W(\by|\bx)\right\}.
\end{equation}
The conditional error event, given $(\bu,\bv,\bx,\by,\calC)$, is
given by
\begin{eqnarray}
\calE(\bu,\bv,\bx,\by,\calC)&=&\bigcup_{\bu'\neq\bu}\left\{
P(\bu',\bv)W(\by|\bx[\bu'])\ge
P(\bu,\bv)W(\by|\bx[\bu])\right\}\nonumber\\
&\dfn&\bigcup_{\bu'\neq\bu}\calE(\bu,\bu',\bv,\bx,\by,\calC)
\end{eqnarray}
The probability of the pairwise error event, 
$\calE(\bu,\bu',\bv,\bx,\by,\calC)$ (again, w.r.t.\ the randomness
of the bin assignment), is given by:
\begin{equation}
\overline{\mbox{Pr}}\{\calE(\bu,\bu',\bv,\bx,\by,\calC)\}=2^{-nR}\bigg|\calC\bigcap
\calA(\bu,\bu',\bv,\bx,\by)\bigg|+2^{-nR}\calI\{P(\bu',\bv)\ge P(\bu,\bv)\}.
\end{equation}
Here, the first term accounts for the probability to randomly choose a bin, other
than $f(\bu)$ , which is
mapped to a channel input vector whose likelihood score is larger than
$P(\bu,\bv)W(\by|\bx[\bu])$.
The second term is associated with the probability that $f(\bu')=f(\bu)$
(which is $2^{-nR}$), in
which case the factor $W(\by|\bx[\bu'])=W(\by|\bx[\bu])$ cancels out in the
pairwise likelihood score comparison, and so, $\bu'$ prevails if
$P(\bu',\bv)\ge P(\bu',\bv)$.
Now, 
\begin{eqnarray}
& &\overline{P}_e(\bu,\bv,\bx,\by,\calC)\nonumber\\
&=&\overline{\mbox{Pr}}\{\calE(\bu,\bv,\bx,\by,\calC)\}\nonumber\\
&\exe&\sum_{\{\calT(\bu'|\bv)\}}
\overline{\mbox{Pr}}\left\{\bigcup_{\bu''\in\calT(\bu'|\bv)}
\calE(\bu,\bu'',\bv,\bx,\by,\calC)\}\right\}\nonumber\\
&\exe&\sum_{\{\calT(\bu'|\bv)\}}\min\left\{1,|\calT(\bu'|\bv)|\cdot 
2^{-nR}\left[\bigg|\calC\bigcap \calA(\bu,\bu',\bv,\bx,\by)\bigg|+\calI\{P(\bu',\bv)\ge
P(\bu,\bv)\}\right]\right\},
\end{eqnarray}
where we have used the fact that $\calA(\bu,\bu'',\bv,\bx,\by)
=\calA(\bu,\bu',\bv,\bx,\by)$ and $P(\bu'',\bv)=P(\bu',\bv)$ for $\bu''\in\calT(\bu'|\bv)$ and 
where the exponential tightness of the truncated union bound (for pairwise
independent events) in the last
expression is known from \cite[Lemma A.2, p.\ 109]{Shulman03} and it can
also be readily deduced from de Caen's lower bound on the probability of a
union of events \cite{deCaen97}.
The next step is to average over the randomness of $\calC$
(except the codeword for the bin of the actual source 
vector $\bu$, which is still given to be $\bx$):
\begin{eqnarray}
\overline{P}_e(\bu,\bv,\bx,\by)&\dfn&\bE_{\calC\setminus\{\bx\}}
\left\{\overline{P}_e(\bu,\bv,\bx,\by,\calC)\right\}\nonumber\\
&\exe&\sum_{\{\calT(\bu'|\bv)\}}\bE\left(\min\left\{1,|\calT(\bu'|\bv)|\cdot 
2^{-nR}\left[\bigg|\calC\bigcap
\calA(\bu,\bu',\bv,\bx,\by)\bigg|+\right.\right.\right.\nonumber\\
& &\left.\left.\left.\calI\{P(\bu',\bv)\ge
P(\bu,\bv)\}\right]\right\}\right),
\end{eqnarray}
where $\bE_{\calC\setminus\{\bx\}}$ stands for expectation over the randomness
of all codewords in $\calC$ other than $\bx=\bx[\bu]$.
Now, using the identity $\bE\{Z\}=\int_0^\infty\mbox{Pr}\{Z\ge t\}\mbox{d}t$,
which is valid for
any non--negative random variable $Z$, we have
\begin{eqnarray}
\label{integral}
& & \bE\left(\min\left\{1,|\calT(\bu'|\bv)|\cdot 
2^{-nR}\left[\bigg|\calC\bigcap
\calA(\bu,\bu',\bv,\bx,\by)\bigg|+\calI\{P(\bu',\bv)\ge
P(\bu,\bv)\}\right]\right\}\right)\nonumber\\
&=&\int_0^1\mbox{d}t\cdot \mbox{Pr}\left\{|\calT(\bu'|\bv)|\cdot
2^{-nR}\left[\bigg|\calC\bigcap
\calA(\bu,\bu',\bv,\bx,\by)\bigg|+\calI\{P(\bu',\bv)\ge
P(\bu,\bv)\}\right]\ge t\right\}\nonumber\\
&=&\int_0^1\mbox{d}t\cdot \mbox{Pr}\left\{
\calI\{P(\bu',\bv)\ge P(\bu,\bv)\}+
\sum_i \calI[\bX(i)\in\calA(\bu,\bu',\bv,\bx,\by)]\ge 
\frac{t\cdot 2^{nR}}{|\calT(\bu'|\bv)|}\right\}\nonumber\\
&\exe&n\ln 2\cdot\int_0^\infty\mbox{d}\theta\cdot 2^{-n\theta}\mbox{Pr}\left\{
\calI\{P(\bu',\bv)\ge P(\bu,\bv)\}+\right.\nonumber\\
& &\left.\sum_i \calI[\bX(i)\in\calA(\bu,\bu',\bv,\bx,\by)]\ge 
2^{n[R-\theta-\hat{H}(U'|V)]}\right\},
\end{eqnarray}
where in the last passage, we have used the shorthand notation $\hat{H}(U'|V)$
for $\hat{H}_{\bu'\bv}(U|V)$, and
we have changed the integration variable from $t$ to
$\theta$, according to the relation $t=2^{-n\theta}$.

Consider first the case where $P(\bu',\bv)\ge P(\bu,\bv)$. Then, the integrand
is given by
\begin{equation}
\label{integrand}
2^{-n\theta}\cdot\mbox{Pr}\left\{1+\sum_i \calI[\bX(i)\in\calA(\bu,\bu',\bv,\bx,\by)]\ge
2^{n[R-\theta-\hat{H}(U'|V)]}\right\}
\end{equation}
in which the second factor is obviously equal to unity for all $\theta \ge [R-\hat{H}(U'|V)]_+$.
Thus, the tail of the integral (\ref{integral}) is given by
\begin{equation}
n\ln 2\cdot\int_{[R-\hat{H}(U'|V)]_+}^\infty\mbox{d}\theta\cdot 2^{-n\theta}
\exe 2^{-n[R-\hat{H}(U'|V)]_+}.
\end{equation}
For $\theta< [R-\hat{H}(U'|V)]_+$, the unity term 
in (\ref{integrand})
can be safely neglected, and 
\begin{equation}
\mbox{Pr}\left\{\sum_i \calI[\bX(i)\in\calA(\bu,\bu',\bv,\bx,\by)]\ge
2^{n[R-\theta-\hat{H}(U'|V)]}\right\}
\end{equation}
is the probability of a large deviations event associated with a binomial
random variable
with $2^{nR}$ trials and probability of success of the exponential order of $2^{-nJ}$, with
$J$ being defined as
\begin{equation}
J\dfn\min\left\{\hat{I}(X';Y):~\hat{P}_{\bx'\by}~\mbox{is such
that}~\bx'\in\calT(Q)\bigcap\calA(\bu,\bu',\bv,\bx,\by)\right\},
\end{equation}
where $\hat{I}(X';Y)$ is shorthand notation for $\hI_{\bx'\by}(X;Y)$
and where it should be kept in mind that $J$ depends on $\hat{P}_{\bu\bv}$,
$\hat{P}_{\bu'\bv}$, and $\hat{P}_{\bx\by}$.
According to \cite[Chap.\ 6]{fnt}, the large deviations
behavior is as follows:
\begin{eqnarray}
& &\mbox{Pr}\left\{\sum_i \calI[\bX(i)\in\calA(\bu,\bu',\bv,\bx,\by)]\ge
2^{n[R-\theta-\hat{H}(U'|V)]}\right\}\nonumber\\
&\exe&\left\{\begin{array}{ll}
2^{-n[J-R]_+} & R-\theta-\hat{H}(U'|V) \le [R-J]_+\\
2^{-n\infty} & R-\theta-\hat{H}(U'|V) > [R-J]_+\end{array}\right.\nonumber\\
&=&\left\{\begin{array}{ll}
2^{-n[J-R]_+} & \theta\ge [R-\hat{H}(U'|V)-[R-J]_+]_+\\
2^{-n\infty} & \mbox{elsewhere}\end{array}\right.
\end{eqnarray}
Thus, the other contribution to (\ref{integral}) is
given by
\begin{eqnarray}
& &n\ln 2\cdot\int_{[R-\hat{H}(U'|V)-[R-J]_+]_+}^{[R-\hat{H}(U'|V)]_+}\mbox{d}\theta\cdot 2^{-n\theta}
\cdot 2^{-n[J-R]_+}\nonumber\\
&\exe& \exp_2\{-n([R-\hat{H}(U'|V)-[R-J]_+]_++[J-R]_+)\}\nonumber\\
&=& \exp_2\{-n([R\wedge J-\hat{H}(U'|V)]_++[J-R]_+)\}\nonumber\\
&=& \exp_2\{-n([R\wedge J-\hat{H}(U'|V)]_+-R\wedge J+R\wedge J+[J-R]_+)\}\nonumber\\
&=& \exp_2\{-n[-R\wedge J\wedge \hat{H}(U'|V)+J]\}\nonumber\\
&=& \exp_2\{-n[J-R\wedge\hat{H}(U'|V)]_+\},
\end{eqnarray}
where we have repeatedly used the identity $a-[a-b]_+=a\wedge b$.
Thus, the total conditional error exponent, for the case $P(\bu',\bv)\ge P(\bu,\bv)$, is 
given by
\begin{eqnarray}
\label{sece}
& &\min\{[R-\hat{H}(U'|V)]_+,[J-R\wedge\hat{H}(U'|V)]_+\}\nonumber\\
&=& \min\{R-R\wedge\hat{H}(U'|V),J-R\wedge J\wedge\hat{H}(U'|V)\}\nonumber\\ 
&=&\left[R\wedge J-\hat{H}(U'|V)\right]_+,
\end{eqnarray}
where the last line follows from the following consideration:\footnote{The
first line of (\ref{sece}) corresponds to the worst between 
the source coding
exponent, $[R-\hat{H}(U'|V)]_+$, and the channel coding exponent,
$[J-R\wedge\hat{H}(U'|V)]_+$, which is to be expected in separate source--
and channel coding.
While the former is non-decreasing in $R$, the
latter is non-increasing. From the last line of (\ref{sece}), we learn that the
overall exponent is non-decreasing in $R$.}
If $\hat{H}(U'|V)>
J$, then all three expressions obviously vanish and then the
equality is trivially met. Otherwise, $\hat{H}(U'|V)\le J$ implies that the term
$R\wedge\hat{H}(U'|V)$, in the second line, can safely be replaced by $R\wedge
J\wedge\hat{H}(U'|V)$, which makes the second line identical to
$R\wedge J-R\wedge J\wedge\hat{H}(U'|V)=\left[R\wedge
J-\hat{H}(U'|V)\right]_+$.
In the case, $P(\bu',\bv)< P(\bu,\bv)$, the conditional error exponent is just
$[J-R\wedge\hat{H}(U'|V)]_+$. 

Let $E_0(\hat{P}_{\bu\bv},\hat{P}_{\bu'\bv},\hat{P}_{\bx\by})$ denote the overall
conditional error exponent given $(\bu,\bu',\bv,\bx,\by)$, i.e.,
\begin{eqnarray}
E_0(\hat{P}_{\bu\bv},\hat{P}_{\bu'\bv},\hat{P}_{\bx\by})
&=&\left\{\begin{array}{ll}
\left[R\wedge J-\hat{H}(U'|V)\right]_+ &
P(\bu',\bv)\ge P(\bu,\bv)\\
\left[J-R\wedge\hat{H}(U'|V)\right]_+ & \mbox{otherwise}\end{array}\right.
\end{eqnarray}
Finally, by averaging the obtained exponential estimate of $\overline{P}_{\mbox{\tiny
e}}(\bU,\bV,\bX,\bY)$ over the randomness of $(\bU,\bV,\bX,\bY)$, and using
the method of types in the standard manner, we obtain
\begin{equation}
E(R,Q)=\lim_{n\to\infty}\min_{\hat{P}_{\bu\bv},\hat{P}_{\bx\by}}
[D(\hat{P}_{\bu\bv}\|P_{UV})+D(\hat{P}_{\by|\bx}\|W|Q)+
E_1(\hat{P}_{\bu\bv},\hat{P}_{\bx\by})],
\end{equation}
where $\hat{P}_{\bx}$ is constrained to coincide with $Q$ and
\begin{equation}
E_1(\hat{P}_{\bu\bv},\hat{P}_{\bx\by})
=\min_{\hat{P}_{\bu'\bv}}E_0(\hat{P}_{\bu\bv},\hat{P}_{\bu'\bv},\hat{P}_{\bx\by}).
\end{equation}
An obvious upper bound\footnote{We are upper bounding the minimum of $E_1$
over $\{\hat{P}_{\bu'\bv}\}$ by the value of $E_0$ where
$\hat{P}_{\bu'\bv}=\hat{P}_{\bu\bv}$, and will shortly see that this bound is
actually tight. This means that the error exponent is dominated by erroneous vectors
$\{\bu'\}$ that are within the same conditional type 
(given $\bv$) as the correct source vector $\bu$. This is coherent with the
observation discussed in Subsection 2.3, that errors within the subcode
pertaining to the same
type class dominate the error exponent.}
is obtained by 
\begin{eqnarray}
E_1(\hat{P}_{\bu\bv},\hat{P}_{\bx\by})
&\le&E_0(\hat{P}_{\bu\bv},\hat{P}_{\bu\bv},\hat{P}_{\bx\by})\nonumber\\
&\le&\left[R\wedge\hat{I}(X;Y)-\hat{H}(U|V)\right]_+\nonumber\\
&\dfn& E_1^*(\hat{P}_{\bu\bv},\hat{P}_{\bx\by}),
\end{eqnarray}
where we have used the fact that $\bx\in\calA(\bu,\bu,\bv,\bx,\by)$ and so,
for $\hat{P}_{\bu'\bv}=\hat{P}_{\bu\bv}$, one has $J\le \hat{I}(X;Y)$.
Thus,
\begin{eqnarray}
E(R,Q)&\le&\min_{P_{U'V'},W'}
\left[D(P_{U'V'}\|P_{UV})+D(W'\|W|Q)
+E_1^*(P_{U'V'},Q\times W')\right]\nonumber\\
&=&\min_{P_{U'V'},W'}\left\{D(P_{U'V'}\|P_{UV})+D(W'\|W|Q)+
\left[R\wedge I(X;Y')-H(U'|V')\right]_+\right\}\nonumber\\
&\dfn& E_U(R,Q).
\end{eqnarray}

We next argue that the universal decoder (\ref{gmmi}) achieves $E_U(R,Q)$
and hence $E(R,Q)=E_U(R,Q)$.
To see why this is true, one repeats exactly the same derivation, with the
following two simple modifications:
\begin{enumerate}
\item $\calA(\bu,\bu',\bv,\bx,\by)$ is replaced by
\begin{equation}
\tilde{\calA}(\bu,\bu',\bv,\bx,\by)=\calT(Q)\cap\{\bx':~\hat{I}(X';Y)-\hH(U'|V)\ge
\hat{I}(X;Y)-\hat{H}(U|V)\} 
\end{equation}
and accordingly, $J$ is replaced by
\begin{equation}
\tilde{J}=\min\{\hat{I}(X';Y):~\hat{I}(X';Y)-\hat{H}(U'|V)\ge \hat{I}(X;Y)-\hat{H}(U|V)\}.
\end{equation}
\item The indicator function $\calI\{P(\bu',\bv)\ge P(\bu,\bv)\}$ is replaced by
$\calI\{\hat{H}(U'|V)\le \hat{H}(U|V)\}$.
\end{enumerate}
The result is then similar except that $E_0(\hat{P}_{\bu\bv},\hat{P}_{\bu'\bv},\hat{P}_{\bx\by})$
is replaced by
\begin{eqnarray}
\label{e0t}
\tilde{E}_0(\hat{P}_{\bu\bv},\hat{P}_{\bu'\bv},\hat{P}_{\bx\by})
&=&\left\{\begin{array}{ll}
[R\wedge\tilde{J}-\hat{H}(U'|V)]_+ &
\hat{H}(U'|V)\le \hat{H}(U|V)\\
\left[\tilde{J}-R\wedge\hat{H}(U'|V)\right]_+ & \mbox{otherwise}\end{array}\right.
\end{eqnarray}
Now, observe that for the first line of
(\ref{e0t}),
\begin{eqnarray}
\label{1stline}
& &R\wedge\tilde{J}-\hat{H}(U'|V)\nonumber\\
&\ge&R\wedge[\hat{I}(X;Y)-\hat{H}(U|V)+\hat{H}(U'|V)]-\hat{H}(U'|V)\nonumber\\
&=&[R-\hat{H}(U'|V)]\wedge[\hat{I}(X;Y)-\hat{H}(U|V)]\nonumber\\
&\ge&[R-\hat{H}(U|V)]\wedge[\hat{I}(X;Y)-\hat{H}(U|V)]
~~~~\mbox{since}~\hat{H}(U'|V)\le
\hat{H}(U|V)\nonumber\\
&=&R\wedge\hat{I}(X;Y)-\hat{H}(U|V),
\end{eqnarray}
where the first line follows from the definition of $\tilde{J}$.
As for the second line,
\begin{eqnarray}
\label{2ndline}
\tilde{J}-R\wedge\hat{H}(U'|V)&\ge&
\hat{I}(X;Y)-\hat{H}(U|V)+\hat{H}(U'|V)-R\wedge\hat{H}(U'|V)\nonumber\\
&=&\hat{I}(X;Y)-\hat{H}(U|V)+[\hat{H}(U'|V)-R]_+\nonumber\\
&\ge&\hat{I}(X;Y)-\hat{H}(U|V)+[\hat{H}(U|V)-R]_+~~~~\mbox{since}~\hat{H}(U'|V)>
\hat{H}(U|V)\nonumber\\
&=&\hat{I}(X;Y)-R\wedge\hat{H}(U|V)\nonumber\\
&\ge&R\wedge\hat{I}(X;Y)-\hat{H}(U|V).
\end{eqnarray}
We conclude then that, no matter whether $\hat{H}(U'|V)\le \hat{H}(U|V)$ or
$\hat{H}(U'|V)>\hat{H}(U|V)$, we always have:
\begin{equation}
\tilde{E}_0(\hat{P}_{\bu\bv},\hat{P}_{\bu'\bv},\hat{P}_{\bx\by})\ge
\left[R\wedge\hat{I}(X;Y)\}-\hat{H}(U|V)\right]_+=E_1^*(\hat{P}_{\bu\bv},\hat{P}_{\bx\by}),
\end{equation}
and so, the overall exponent $E_U(R)$ is achieved by (\ref{gmmi}).

As for the alternative universal decoding metric (\ref{gmmi2}),
the derivation is, once again, the very same, with the pairwise error event
$\calA(\bu,\bu',\bv,\bx,\by)$ and the variable $\tilde{J}$ redefined
accordingly as the minimum of $\hat{I}(X';Y)$ s.t.\
$R\wedge\hat{I}(X';Y)-\hat{H}(U'|V)\ge 
R\wedge\hat{I}(X;Y)-\hat{H}(U|V)$, and then the last two inequalities are
modified as follows:
Instead of (\ref{1stline}), we have
\begin{eqnarray}
& &R\wedge\tilde{J}-\hat{H}(U'|V)\nonumber\\
&=&R\wedge R\wedge\hat{I}(X';Y)-\hat{H}(U'|V)\nonumber\\
&\ge&R\wedge[R\wedge\hat{I}(X;Y)-\hat{H}(U|V)+\hat{H}(U'|V)]-\hat{H}(U'|V)\nonumber\\
&=&[R-\hat{H}(U'|V)]\wedge\{[R\wedge\hat{I}(X;Y)]-\hat{H}(U|V)\}\nonumber\\
&\ge&[R-\hat{H}(U|V)]\wedge\{[R\wedge\hat{I}(X;Y)]-\hat{H}(U|V)\}
~~~~\mbox{since}~\hat{H}(U'|V)\le
\hat{H}(U|V)\nonumber\\
&=&R\wedge\hat{I}(X;Y)-\hat{H}(U|V).
\end{eqnarray}
and instead of (\ref{2ndline}):
\begin{eqnarray}
\tilde{J}-R\wedge\hat{H}(U'|V)&\ge&
R\wedge\tilde{J}-R\wedge\hat{H}(U'|V)\nonumber\\
&\ge&
R\wedge\hat{I}(X;Y)-\hat{H}(U|V)+\hat{H}(U'|V)-R\wedge\hat{H}(U'|V)\nonumber\\
&=&R\wedge\hat{I}(X;Y)-\hat{H}(U|V)+[\hat{H}(U'|V)-R]_+\nonumber\\
&\ge&R\wedge\hat{I}(X;Y)-\hat{H}(U|V)+[\hat{H}(U|V)-R]_+~~~~\mbox{since}~\hat{H}(U'|V)>
\hat{H}(U|V)\nonumber\\
&=&R\wedge\hat{I}(X;Y)-R\wedge\hat{H}(U|V)\nonumber\\
&\ge&R\wedge\hat{I}(X;Y)-\hat{H}(U|V).
\end{eqnarray}
This completes the proof of Theorem 1.

\section{Extensions}

As mentioned in the Introduction, in this section, we provide extensions of the above
results in several directions, including: (i) finite--state sources and
channels with LZ universal decoding 
metrics, (ii) arbitrary sources and channels with universal decoding w.r.t.\ a
given class of metric decoders, and (iii) separate source--channel encodings
and joint universal decoding of correlated source. 
While in (i) and (ii) we no longer expect to have single--letter
formulae for the error exponent, we will still be able to propose
asymptotically optimum universal decoding metrics in the error exponent sense.
While the skeleton of the analysis builds upon the one
of the proof of Theorem 1, we
will highlight the non--trivial differences and the modifications needed relative to 
the proof of Theorem 1.

\subsection{Finite--State Sources/Channels and a Universal LZ Decoding Metric}

In \cite{Ziv85}, Ziv considered the class of finite--state channels and 
proposed a universal decoding metric that is based on conditional LZ parsing.
Here, we discuss a similar model with a suitable extension of Ziv's decoding metric in
the spirit of the generalized MMI decoder.

Consider a sequence of pairs of random variables $\{(U_i,V_i)\}_{i=1}^n$, drawn from a
finite--alphabet, finite--state source, defined according to
\begin{equation}
\label{fss}
P(\bu,\bv)=\prod_{t=1}^n
P(u_t,v_t|s_t)
\end{equation}
where $s_t$ is the joint state of the two sources at time $t$, which evolves
according to 
\begin{equation}
s_t=g(s_{t-1},u_{t-1},v_{t-1}),
\end{equation}
with $g:\calS\times\calU\times\calV\to\calS$ being the source next--state
function, and $\calS$ being a finite set of
states. The initial state, $s_1$, is assumed to be an arbitrary fixed member of
$\calS$. By the same
token, the channel is also assumed to be finite--state (as
in \cite{Ziv85}), i.e.,
\begin{equation}
\label{fsc}
W(\by|\bx)=\prod_{t=1}^n
W(y_t|x_t,z_t),~~~~~z_t=h(z_{t-1},x_{t-1},y_{t-1}),
\end{equation}
where $z_t$ is the channel state at time $t$, taking on values in a finite set
$\calZ$ and $h:\calZ\times\calX\times\calY\to\calZ$ is the channel next--state function.
Once again, the initial state, $z_1$, is an arbitrary member of $\calZ$.

The remaining details of the communication system are the same as described in
Subsection 2.2, with the exception that the random coding distribution, now
denoted by $Q(\bx)$, is
allowed here to be more general than a uniform distribution across a type
class (or the uniform distribution across $\calX^n$, as assumed in
\cite{Ziv85}). Similarly as in \cite{givenclassofmetrics}, we
assume that $Q$ may be any exchangeable probability distribution
(i.e., $\bx'$ is a permutation of $\bx$ implies $Q(\bx')=Q(\bx)$), and that, moreover,
if the state variable $z_t$ includes a component, say, $\sigma_t$, that
is fed merely by $\{x_t\}$ (but not $\{y_t\}$),
then it is enough that $Q$ would be invariant within
conditional types of $\bx$ given $\bsigma=(\sigma_1,\ldots,\sigma_n)$.

Let $\hH_{\mbox{\tiny LZ}}(\bx|\by)$ denote the normalized conditional LZ
compressibility of $\bx$ given $\by$, as defined in \cite[eq.\ (20)]{Ziv85}
(and denoted by $u(\bx,\by)$ therein).\footnote{This means that
$n\hH_{\mbox{\tiny LZ}}(\bx|\by)$ is the length of the conditional Lempel--Ziv
code for $\bx$, where $\by$ serves as SI available to both the encoder and
decoder, which is based on joint incremental parsing of the sequence pair 
$(\bx,\by)$ (see also \cite{fsmessages}).
Here, we are deliberately
using a somewhat different notation than the usual, which hopefully makes the analogy to the memoryless
case self--evident.} Next define
\begin{equation}
\label{lzmi}
\hI_{\mbox{\tiny LZ}}(\bx;\by)=-\frac{\log Q(\bx)}{n}-\hH_{\mbox{\tiny
LZ}}(\bx|\by),
\end{equation}
and finally, define the universal decoder 
\begin{equation}
\label{lzdecoder}
\tilde{\bu}=\arg\max_{\bu}\left[\hI_{\mbox{\tiny LZ}}(\bx[\bu];\by)-\hH_{\mbox{\tiny
LZ}}(\bu|\bv)\right].
\end{equation}
Note that the first term on the r.h.s.\ of (\ref{lzmi}) plays a role analogous
to that of the unconditional empirical entropy, $\hH_{\bx}(X)$, of the memoryless case
(and indeed, at least for the uniform distribution over a type class, as assumed in the
previous sections, it is asymptotically equivalent), and so, the difference
in (\ref{lzmi}) makes sense as an extension of the empirical mutual information between $\bx$
and $\by$.

As in Theorem 1, part b (and as an extension to \cite{Ziv85}), 
we now argue that the universal decoder (\ref{lzdecoder}) 
achieves an average error probability that is, within a sub-exponential 
function of $n$, the same as the average error probability of the MAP decoder
for the given source (\ref{fss}) and channel (\ref{fsc}).

\begin{theorem}
Consider the problem setting defined in Subsection 2.2, with a finite--state
source (\ref{fss}) and a finite--state channel (\ref{fsc}). Assume that the random
binning ensemble is as before and that the random channel coding distribution $Q$ 
is as described in the third paragraph of this subsection. Let
$\overline{P}_{\mbox{\tiny e}}^{\mbox{\tiny MAP}}(n)$ denote the average error probability
of the MAP decoder and let
$\overline{P}_{\mbox{\tiny e}}^{\mbox{\tiny u}}(n)$ denote the average error probability
of the decoder (\ref{lzdecoder}). Then,
\begin{equation}
\lim_{n\to\infty}\frac{1}{n}\log\frac{
\overline{P}_{\mbox{\tiny e}}^{\mbox{\tiny u}}(n)}{
\overline{P}_{\mbox{\tiny e}}^{\mbox{\tiny MAP}}(n)}=0.
\end{equation}
\end{theorem}
In other words, similarly as in \cite{Ziv85}, while we do not have a
characterization of the error exponent, we can still guarantee that
whenever the MAP decoder has an exponentially decaying average error
probability, then so does the decoder (\ref{lzdecoder}), and with the same exponential rate.

\vspace{0.2cm}

\noindent
{\it Proof outline.}
The skeleton of the proof of Theorem 2 is similar to the proof of Theorem 1,
but as mentioned earlier, some non--trivial modifications are needed.
Below we outline the main steps, highlighting the main modifications required.

\vspace{0.1cm}

\noindent
1. The conditional type class of $\bu$ given $\bv$, $\calT(\bu|\bv)$, is redefined as
the set $\{\bu':~P(\bu',\bv)=P(\bu,\bv)\}$, where $P$ is given as in
(\ref{fss}).\footnote{Note that the requirement $P(\bu',\bv)=P(\bu,\bv)$ is
imposed here only for the given $P$, not even for every finite--state source in the
class.} Obviously, for every given $\bv$, the various `types'
$\{\calT(\bu|\bv)\}$ are equivalence
classes, and hence form a partition of $\calU^n$. 
One important property that would essential for the proof is that the number
$K_n(\bv)$ of distinct types,
$\{\calT(\bu|\bv)\}$, under this definition, for a given $\bv$, grows sub--exponentially
in $n$ (just like in the case of ordinary types). 
This guarantees that the probability of a union of events, over $\{\calT(\bu'|\bv)\}$,
is of the same exponential order as the maximum term, as was the case in the
proof of Theorem 1.
Interestingly, this can
easily be proved using the theory of LZ data compression: 
\begin{equation}
K_n(\bv)=\sum_{\bu\in\calU^n}\frac{1}{|\calT(\bu|\bv)|}
\le\sum_{\bu\in\calU^n}2^{-n\hH_{\mbox{\tiny LZ}}(\bu|\bv)+o(n)}\le 2^{o(n)},
\end{equation}
where $o(n)$ stands for a sub-linear term (i.e., $\lim_{n\to\infty} o(n)/n =0$, 
uniformly in both $\bu$ and $\bv$),
the first inequality is by \cite[Lemma 1, p.\ 459]{Ziv85}\footnote{Not to be
confused with the lemma on page 456 of \cite{Ziv85}, which is also referred to
as Lemma 1.} and the second
inequality is due to the fact that $n\hH_{\mbox{\tiny LZ}}(\bu|\bv)$ is
(within negligible terms) a legitimate length function for lossless
compression of $\bu$ (with SI $\bv$) (see \cite[Lemma 2]{Ziv85} and
\cite{fsmessages}) and
hence must satisfy the Kraft inequality for every given $\bv$.

\vspace{0.1cm}

\noindent
2. The quantity $\hH(U'|V)$, in the proof of Theorem 1, is replaced by
$\frac{1}{n}\log|\calT(\bu'|\bv)|$ with the above modified definition 
of the conditional type.

\vspace{0.1cm}

\noindent
3. The definition of $J$ is changed to
\begin{equation}
J=\min\left\{-\frac{1}{n}\log Q[\calT(\bx'|\by)]:
~\bx'\in\calA(\bu,\bu',\bv,\bx,\by)\right\},
\end{equation}
where $\calA(\bu,\bu',\bv,\bx,\by)$ is the pairwise error event pertaining to the
MAP decoder (for the lower bound) or to the universal decoder
(\ref{lzdecoder}) (for the upper bound). By our assumptions, $Q$ assigns the
same probability to all members of $\calT(\bx'|\by)$, thus 
\begin{equation}
\frac{1}{n}\log Q[\calT(\bx'|\by)]=\frac{\log
Q(\bx')}{n}+\frac{\log|\calT(\bx'|\by)|}{n}.
\end{equation}

\vspace{0.1cm}

\noindent
4. Using the above, and following the same steps as in the proof of Theorem
1, the conditional average error probability, 
given $(\bu,\bv,\bx,\by)$, associated with the MAP decoder, can be shown to be lower
bounded by an expression of the exponential order of
$\exp\{-nE_0(\bu,\bv,\bx,\by)\}$, 
where
\begin{eqnarray}
E_0(\bu,\bv,\bx,\by)
&\le&\left[\min\left\{R,-\frac{1}{n}\log
Q[\calT(\bx|\by)]\right\}-\frac{1}{n}\log|\calT(\bu|\bv)|\right]_+\nonumber\\
&=&\left[\min\left\{R,-\frac{1}{n}\log Q(\bx)-\frac{1}{n}\log
|\calT(\bx|\by)|\right\}-\frac{1}{n}\log|\calT(\bu|\bv)|\right]_+\nonumber\\
&\le&\left[\min\left\{R,-\frac{1}{n}\log Q(\bx)-\hH_{\mbox{\tiny LZ}}(\bx|\by)
\right\}-\hH_{\mbox{\tiny LZ}}(\bu|\bv)\right]_++o(1)\nonumber\\
&=&\left[R\wedge\hI_{\mbox{\tiny LZ}}(\bx;\by)
-\hH_{\mbox{\tiny LZ}}(\bu|\bv)\right]_++o(1)\nonumber\\
&\dfn& E_1^*(\bu,\bv,\bx,\by),
\end{eqnarray}
and where we have used twice Lemma 1 of \cite[p.\ 459]{Ziv85} and
the fact that $\bx\in\calA(\bu,\bu,\bv,\bx,\by)$ and so,
for $P(\bu',\bv)=P(\bu,\bv)$, one has $J\le -\frac{1}{n}\log Q[\calT(\bx|\by)]$.

\vspace{0.1cm}

\noindent
5. For the upper bound on the error probability of (\ref{lzdecoder}), 
$\calA(\bu,\bu',\bv,\bx,\by)$ is replaced by
\begin{equation}
\tilde{\calA}(\bu,\bu',\bv,\bx,\by)=\{\bx':~I_{\mbox{\tiny
LZ}}(\bx';\by)-\hH_{\mbox{\tiny LZ}}(\bu'|\bv)\ge
I_{\mbox{\tiny LZ}}(\bx;\by)-\hH_{\mbox{\tiny LZ}}(\bu|\bv)\}
\end{equation}
and accordingly, $J$ is replaced by
\begin{eqnarray}
\tilde{J}&=&\min\{I_{\mbox{\tiny LZ}}(\bx';\by):~I_{\mbox{\tiny
LZ}}(\bx';\by)-\hH_{\mbox{\tiny LZ}}(\bu'|\bv)\ge 
I_{\mbox{\tiny LZ}}(\bx;\by)-\hH_{\mbox{\tiny LZ}}(\bu|\bv)\}\nonumber\\
&=&I_{\mbox{\tiny LZ}}(\bx;\by)-\hH_{\mbox{\tiny
LZ}}(\bu|\bv)+\hH_{\mbox{\tiny LZ}}(\bu'|\bv).
\end{eqnarray}

\vspace{0.1cm}

\noindent
6. The indicator function $\calI[P(\bu',\bv)\ge P(\bu,\bv)]$ is replaced by
$\calI[\hH_{\mbox{\tiny LZ}}(\bu'|\bv)\le \hH_{\mbox{\tiny LZ}}(\bu|\bv)]$.

\vspace{0.1cm}

\noindent
7. For the error probability analysis of the universal decoder
(\ref{lzdecoder}), the union over erroneous source vectors $\{\bu'\}$ is
partitioned into (a sub--exponential number of) `types' of the form
\begin{equation}
\calT_\ell(\bu'|\bv)=\{\tilde{\bu}:~P(\tilde{\bu},\bv)=
P(\bu',\bv),~n\hH_{\mbox{\tiny
LZ}}(\tilde{\bu}|\bv)=\ell\},
\end{equation}
for $\ell=1,2,\ldots$, and one uses the fact that
$|\calT_\ell(\bu'|\bv)|\le 2^\ell$, as $n\hH_{\mbox{\tiny LZ}}(\cdot|\bv)$
is a length function of a lossless data compression algorithm.

\vspace{0.1cm}

\noindent
8. It is observed that
$\sum_i \calI[\bX(i)\in\tilde{\calA}(\bu,\bu',\bv,\bx,\by)]$ 
is a binomial random variables
with $2^{nR}$ trials and probability of success of the exponential order of $2^{-n\tilde{J}}$.
To see why the latter is true, consider the following:
\begin{eqnarray}
& &Q\left\{\bX'\in\tilde{\calA}(\bu,\bu',\bv,\bx,\by)\right\}\nonumber\\
&=& Q\left\{I_{\mbox{\tiny LZ}}(\bX';\by)\ge 
I_{\mbox{\tiny LZ}}(\bx;\by)-\hH_{\mbox{\tiny LZ}}(\bu|\bv)+\hH_{\mbox{\tiny
LZ}}(\bu'|\bv)\right\}\nonumber\\
&=&\sum_{\{\bx':~I_{\mbox{\tiny LZ}}(\bx';\by)\ge 
I_{\mbox{\tiny LZ}}(\bx;\by)-\hH_{\mbox{\tiny LZ}}(\bu|\bv)+\hH_{\mbox{\tiny
LZ}}(\bu'|\bv)\}} Q(\bx')\nonumber\\
&=&\sum_{\{\bx':~I_{\mbox{\tiny LZ}}(\bx';\by)\ge 
I_{\mbox{\tiny LZ}}(\bx;\by)-\hH_{\mbox{\tiny LZ}}(\bu|\bv)+\hH_{\mbox{\tiny
LZ}}(\bu'|\bv)\}}
Q(\bx')2^{n\hH_{\mbox{\tiny LZ}}(\bx'|\by)}\cdot 2^{-n\hH_{\mbox{\tiny LZ}}(\bx'|\by)}\nonumber\\
&=&\sum_{\{\bx':~I_{\mbox{\tiny LZ}}(\bx';\by)\ge 
I_{\mbox{\tiny LZ}}(\bx;\by)-\hH_{\mbox{\tiny LZ}}(\bu|\bv)+\hH_{\mbox{\tiny
LZ}}(\bu'|\bv)\}}
\exp_2\{-nI_{\mbox{\tiny LZ}}(\bx';\by)\}\cdot 2^{-n\hH_{\mbox{\tiny LZ}}(\bx'|\by)}\nonumber\\
&\le&\sum_{\bx'\in\calX^n} 
\exp_2\{-n[I_{\mbox{\tiny LZ}}(\bx;\by)-\hH_{\mbox{\tiny
LZ}}(\bu|\bv)+\hH_{\mbox{\tiny LZ}}(\bu'|\bv)]\}
\cdot 2^{-n\hH_{\mbox{\tiny LZ}}(\bx'|\by)}\nonumber\\
&\le&
\exp_2\{-n[I_{\mbox{\tiny LZ}}(\bx;\by)-\hH_{\mbox{\tiny
LZ}}(\bu|\bv)+\hH_{\mbox{\tiny LZ}}(\bu'|\bv)]\}
\sum_{\bx'\in\calX^n} 
2^{-n\hH_{\mbox{\tiny LZ}}(\bx'|\by)}\nonumber\\
&\le&
\exp_2\{-n[I_{\mbox{\tiny LZ}}(\bx;\by)-\hH_{\mbox{\tiny
LZ}}(\bu|\bv)+\hH_{\mbox{\tiny LZ}}(\bu'|\bv)]+o(n)\},
\end{eqnarray}
where in the last step, we have used again Kraft's inequality.

\vspace{0.1cm}

\noindent
9. Using exactly the same method as in the proof of Theorem 1, one can show that that conditional error
probability of the universal decoder (\ref{lzdecoder}) is upper bounded by an
expression whose exponential order is lower bounded by
$E_1^*(\bu,\bv,\bx,\by)$.

It should be noted that these results continue to apply for
{\it arbitrary} sources and channels (even deterministic ones), where the assertion
would be that the decoder
(\ref{lzdecoder}) competes favorably (in the error exponent sense) relative to
any decoding metric of the form 
\begin{equation}
\sum_{t=1}^nm_{\mbox{\tiny s}}(u_t,v_t,s_t)+
\sum_{t=1}^nm_{\mbox{\tiny c}}(x_t,y_t,z_t), 
\end{equation}
where $s_t$ and $z_t$ evolve according to
next--state functions $h$ and $g$, as defined above.
This follows from the observation that the assumption on underlying finite--state sources and
finite--state channels was actually used merely in the assumed structure of the MAP decoding
metric, with which decoder (\ref{lzdecoder}) competes. The fact that the
overall probability of error is eventually averaged over all source vectors and
channel noise realizations pertaining to finite--state probability
distributions, was not really used here, since we compared the conditional error
probabilities given $(\bu,\bv,\bx,\by)$. The same observation has been exploited also
in \cite{givenclassofmetrics} for universal pure 
channel coding, and it will be further developed in the next
subsection.

\subsection{Arbitrary Sources and Channels With a Given Class of Metric Decoders}

In \cite{givenclassofmetrics}, the following setting of universal channel
decoding was studied: Given a random coding distribution $Q$, for independent
random selection of $2^{nR}$ codewords $\{\bx_i\}$, and given a limited class of reference
decoders, defined by a family of decoding metrics $\{m_\theta(\bx,\by),~\theta\in\Theta\}$
($\theta$ being an index or a parameter), find a decoding metric that is universal in
the sense of achieving an average error probability that is, within a
sub--exponential function of $n$, as good as the best decoder in the class, no
matter what the underlying
channel, $W(\by|\bx)$, may be. The following decoder was shown in
\cite{givenclassofmetrics} to possess this property under a certain condition
that will be specified shortly:
estimate the message $i$ as the one that minimizes $u(\bx_i,\by)=\log Q[\calT(\bx_i|\by)]$,
where $\calT(\bx|\by)$ designates a notion of a ``type'' induced by the family 
of decoding metrics (rather than by channels), namely, 
\begin{equation}
\calT(\bx|\by)=\left\{\bx':~m_\theta(\bx',\by)=m_\theta(\bx,\by)~\forall~\theta\in\Theta\right\}.
\end{equation}
As $\{\calT(\bx|\by)\}$ are equivalence classes, they form a partition of
$\calX^n$ for every given $\by$. The condition required for the universality
of this decoding metric is that the number of distinct `types' $\{\calT(\bx|\by)\}$
would grow sub--exponentially with $n$.

A similar approach can be taken in the present problem setting. Given a family
of decoding metrics of the form\footnote{This additive structure can be
justified by the fact that the MAP decoding metric is also additive,
as it maximizes $\log P(\bu,\bv)+\log W(\by|\bx[\bu])$.}
\begin{equation}
m_\theta(\bu,\bv,\bx,\by)=m_{\mbox{\tiny s},\theta}(\bu,\bv)+
m_{\mbox{\tiny c},\theta}(\bx,\by),~~~~\theta\in\Theta,
\end{equation}
let us define
\begin{equation}
\calT_{\mbox{\tiny s}}(\bu|\bv)=\{\bu':~m_{\mbox{\tiny
s},\theta}(\bu',\bv)=m_{\mbox{\tiny
s},\theta}(\bu,\bv)~\forall~\theta\in\Theta\}
\end{equation}
\begin{equation}
\calT_{\mbox{\tiny s}}(\bx|\by)=\{\bx':~m_{\mbox{\tiny
c},\theta}(\bx',\by)=m_{\mbox{\tiny
c},\theta}(\bx,\by)~\forall~\theta\in\Theta\},
\end{equation}
and assume, as before, that the numbers of distinct `types',
$\{\calT_{\mbox{\tiny s}}(\bu|\bv)\}$ and 
$\{\calT_{\mbox{\tiny c}}(\bx|\by)\}$, both grow sub--exponentially with $n$.
Then, the universal decoder
\begin{equation}
\hat{\bu}=\arg\min_{\bu}\left\{\log|\calT_{\mbox{\tiny s}}(\bu|\bv)|+
\log Q[\calT_{\mbox{\tiny c}}(\bx[\bu]|\by)]\right\}
\end{equation}
competes favorably with all metrics in the above family, no matter what the
underlying source and the underlying channel may be. The proof combines the
ideas of the proof of
Theorem 1 above with those of \cite{givenclassofmetrics}, with the
proper adjustments, of course, but it is otherwise straightforward. 
Here, the term  $\log|\calT_{\mbox{\tiny s}}(\bu|\bv)|$
is the analogue of $n$ times the conditional empirical entropy pertaining to the source
part, whereas the term
$\log Q[\calT_{\mbox{\tiny c}}(\bx[\bu]|\by)$ plays the role of
$n$ times the negative empirical mutual information between $\bx[\bu]$ and $\by$.
Therefore if, for example, 
\begin{eqnarray}
m_{\mbox{\tiny c},\theta}(\bx,\by)&=&\sum_{t=1}^n
m_{\mbox{\tiny c},\theta}(x_t,y_t)~~~\mbox{and}\\
m_{\mbox{\tiny s},\theta}(\bu,\bv)&=&\sum_{t=1}^n
m_{\mbox{\tiny s},\theta}(u_t,v_t), 
\end{eqnarray}
as is the case when the sources and the
channel are memoryless, then $\{\calT_{\mbox{\tiny c}}(\bx|\by)\}$
and $\{\calT_{\mbox{\tiny s}}(\bu|\bv)\}$ become conditional type classes in the
usual sense, and 
we are back to the generalized MMI decoder of
Section 3, provided that $Q$ is, again, the uniform distribution within a single type class. 
As a final note, in this context, we mention that
in this setting, the input and the output alphabets of the channel may also be
continuous, see, e.g., \cite[p.\ 5575, Example 3]{givenclassofmetrics}.

\subsection{Separate Encodings and Universal Joint Decoding of Correlated Sources}

\begin{figure}[ht]
\hspace*{1cm}\input{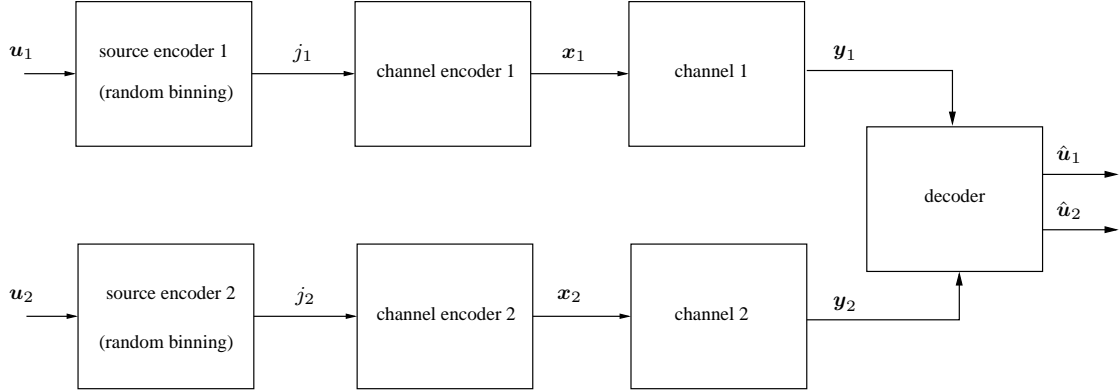}
\caption{\small Separate source--channel encodings and joint decoding of two correlated sources.}
\label{sw2}
\end{figure}

Consider the system depicted in Fig.\ \ref{sw2}, which illustrates a scenario
of separate source--channel encodings and joint decoding of two correlated
sources, $\bu_1$ and $\bu_2$. For the sake of simplicity of the presentation,
we return to the assumption of memoryless systems, as in Section 3.

Consider $n$ independent copies $\{(U_{1,i},U_{2,i})\}_{i=1}^n$ of a finite--alphabet
pair of random variables $(U_1,U_2)\sim
P_{U_1U_2}$, as well as $n$ uses of two independent finite--alphabet DMC's
$W_1(\by_1|\bx_1)=\prod_{t=1}^n
W_1(y_{1,t}|x_{1,t})$ and
$W_2(\by_2|\bx_2)=\prod_{t=1}^n
W_2(y_{2,t}|x_{2,t})$.
For $k=1,2$, consider the following mechanism:
The source vector $\bu_k=(u_{k,1},\ldots,u_{k,n})$ 
is encoded into one out of
$M_k=2^{nR_k}$
bins, selected independently at random for every member of $\calU_k^n$.
The bin index $j_k=f_k(\bu_k)$ is in turn mapped into a channel input vector
$\bx_k(i)\in\calX_1^n$, which is transmitted across the channel $W_k$.
The various codewords $\{\bx_k(i)\}_{i=1}^{M_k}$
are selected independently at
random under the uniform distribution within given type classes $\calT(Q_k)$,
where $Q_k$ is a given distribution across $\calX_k$.
The randomly chosen codebook
$\{\bx_k(1),\bx_k(2),\ldots,\bx_k(M_k)\}$
will be denoted by $\calC_k$. Similarly, as before,
we will sometimes denote $\bx_k(j_k)=\bx_k[f_k(\bu_k)]$
by $\bx_k[\bu_k]$.
The optimal (MAP) decoder estimates $(\bu_1,\bu_2)$, using the channel
outputs $\by_1$ and $\by_2$, according to
\begin{equation}
\label{mapdecoder2}
(\hat{\bu}_1,\hat{\bu}_2)=
\arg\max_{\bu_1,\bu_2}
P(\bu_1,\bu_2)W_1(\by_1|\bx_1[\bu_1])
W_2(\by_2|\bx_2[\bu_2]).
\end{equation}

The main structure of the analysis continues to be essentially the same as in Section 4.
The situation here, however, is significantly more involved, because five
different types of pairwise error events $\{(\bu_1,\bu_2)\to(\bu_1',\bu_2')\}$ 
should be carefully handled:
\begin{enumerate}
\item $\bu_1'\ne\bu_1$ and $\bu_2'=\bu_2$.
\item $\bu_2'\ne\bu_2$ and $\bu_1'=\bu_1$.
\item Both $\bu_1'\ne\bu_1$ and $\bu_2'\ne\bu_2$, but (at least)
$\bu_2'$ is
mapped into the same bin as $\bu_2$.
\item Both $\bu_1'\ne\bu_1$ and $\bu_2'\ne\bu_2$, but (at
least)\footnote{Here, we are counting twice the case ``$\bu_1'\ne\bu_1$ and
$\bu_2'\ne\bu_2$ and both estimates are in the bins of their respective
true source vectors.'' This is done simply for symmetry the structure above, without affecting the
error exponent.}
$\bu_1'$ is
mapped into the same bin as $\bu_1$.
\item Both $\bu_1'\ne\bu_1$ and $\bu_2'\ne\bu_2$, and neither $\bu_1'$ nor $\bu_2'$
belongs to the same bin as the respective true source vector.
\end{enumerate}
Errors of types 1 and 2 are of the same nature as in Section 3, where the
source that is estimated correctly, is actually in the role of SI at the decoder. 
Following (\ref{gmmi2}), the respective metrics are\footnote{Note that $f_1$
does not really depend on $(\bx_2,\by_2)$, and similarly, 
$f_2$ does not depend on $(\bx_1,\by_1)$. Nonetheless, we deliberately adopt
this uniform notation for convenience later on.}
\begin{eqnarray}
f_1(\bu_1,\bu_2,\bx_1,\bx_2,\by_1,\by_2)&=&R_1\wedge
\hI(X_1;Y_1)-\hH(U_1|U_2)\\
f_2(\bu_1,\bu_2,\bx_1,\bx_2,\by_1,\by_2)&=&R_2\wedge
\hI(X_2;Y_2)-\hH(U_2|U_1).
\end{eqnarray}
where $\hI(X_1;Y_1)$ and $\hH(U_1|U_2)$ are shorthand notations for
$\hI_{\bx_1\bx_2}(X_1;Y_1)$ and $\hH_{\bu_1\bu_2}(U_1|U_2)$, respectively, and
so on. Errors of types 3 and 4 will turn out to be addressed by metrics of the form
\begin{eqnarray}
f_3(\bu_1,\bu_2,\bx_1,\bx_2,\by_1,\by_2)&=&R_1\wedge
\hI(X_1;Y_1)+R_2-\hH(U_1,U_2)\\
f_4(\bu_1,\bu_2,\bx_1,\bx_2,\by_1,\by_2)&=&R_2\wedge
\hI(X_2;Y_2)+R_1-\hH(U_1,U_2).
\end{eqnarray}
Finally, error of type 5 is accommodated by
\begin{eqnarray}
f_5(\bu_1,\bu_2,\bx_1,\bx_2,\by_1,\by_2)&=&\hI(X_1;Y_1)+\hI(X_2;Y_2)-\nonumber\\
& &\min\{\hH(U_1,U_2),
R_1\wedge\hI(X_1;Y_1)+R_2\wedge\hI(X_2;Y_2)\}\nonumber\\
&\equiv&
[R_1\wedge\hI(X_1;Y_1)+R_2\wedge\hI(X_2;Y_2)-\hH(U_1,U_2)]_++\nonumber\\
& &[\hI(X_1;Y_1)-R_1]_++[\hI(X_2;Y_2)-R_2]_+
\end{eqnarray}
But we need a {\it single} universal decoding
metric that copes with all five types of errors at the same time.

Similarly as in \cite[eqs.\ (57)-(60)]{givenclassofmetrics}, this objective
is accomplished by a metric which is given by the minimum among all five
metrics above, i.e., we define
our decoding metric as
\begin{equation}
f_0(\bu_1,\bu_2,\bx_1,\bx_2,\by_1,\by_2)=\min_{1\le i\le 5}
f_i(\bu_1,\bu_2,\bx_1,\bx_2,\by_1,\by_2),
\end{equation}
Our main result in this subsection is the following.

\begin{theorem}
Consider the above described setting of separate encodings and joint decoding
of two correlated memoryless sources transmitted over two respective,
independent memoryless channels. Then, the universal decoder
\begin{equation}
(\tilde{\bu}_1,\tilde{\bu}_2)=\mbox{arg}\max_{\bu_1,\bu_2}
f_0(\bu_1,\bu_2,\bx_1[\bu_1],\bx_2[\bu_2],\by_1,\by_2) 
\end{equation}
achieves the same
random--binning/random--coding error exponent as the MAP decoder
(\ref{mapdecoder2}).
\end{theorem}

{\it Proof outline.}
The conditional probability of error given
$(\bu_1,\bu_2,\bx_1,\bx_2,\by_1,\by_2)$, for both
the MAP decoder and the universal decoder, can be shown to be of the exponential order of 
$$\exp_2\{-n[f_0(\bu_1,\bu_2,\bx_1,\bx_2,\by_1,\by_2)]_+\}.$$
To show this, the analysis of the probability of error, for both the MAP decoder and the
universal decoder, should be divided into several parts, according to the
various types of error events. Errors of types 1 and
2 are addressed exactly as in Section 4.
The more complicated part of the analysis is due to errors of types 3--5,
where both competing source vectors are in error. However,
this analysis too follows the same basic ideas. Here we will outline only
the main ingredients that are different from those of the proof of Theorem 1.

For a given $\bu_1'\ne\bu_1$ and $\bu_2'\ne \bu_2$ (errors of types 3--5), let us define
the pairwise error event
\begin{eqnarray}
& &\calA(\bu_1,\bu_1',\bu_2,\bu_2',\bx_1,\bx_2,\by_1,\by_2)\nonumber\\
&=&[\calT(Q_1)\times\calT(Q_2)]\bigcap\nonumber\\
& &\left\{(\bx_1',\bx_2'):~P(\bu_1',\bu_2')W_1
(\by_1|\bx_1')W_2(\by_2|\bx_2')\ge
P(\bu_1,\bu_2)W_1
(\by_1|\bx_1)W_2
(\by_2|\bx_2)\right\}.\nonumber
\end{eqnarray}
The conditional error event, given
$(\bu_1,\bu_2,\bx_1,\bx_2,\by_1,\by_2,\calC_1,\calC_2)$, is
given by
\begin{eqnarray}
& &\calE(\bu_1,\bu_2,\bx_1,\bx_2,\by_1,\by_2,\calC_1,\calC_2)\nonumber\\
&=&\bigcup_{\bu_1'\neq\bu_1,~\bu_2'\ne\bu_2}\left\{
P(\bu_1',\bu_2')W_1(\by_1|\bx_1[\bu_1'])
W_2(\by_2|\bx_2[\bu_2'])\ge\right.\nonumber\\
& &\left.P(\bu_1,\bu_2)W_1(\by_1|\bx_1[\bu_1])W_2(\by_2|\bx_2[\bu_2])
\right\}\nonumber\\
&\dfn&
\bigcup_{\bu_1'\ne\bu_1,~\bu_2'\ne\bu_2}\calE(\bu_1,\bu_1',\bu_2,\bu_2',\bx_1,\bx_2,\by_1,\by_2,
\calC_1,\calC_2)
\end{eqnarray}
Here too, the exponential tightness of the truncated union bound for two dimensional
unions of events with independence structure as above can be established 
using de Caen's lower bound \cite{deCaen97} (see \cite{SMF13}).
For errors of types 3 and 4, let us define
\begin{eqnarray}
& &\calA_1(\bu_1,\bu_1',\bu_2,\bu_2',\bx_1,\by_1)\nonumber\\
&=&\calT(Q_1)\bigcap
\left\{\bx_1':~P(\bu_1',\bu_2')W_1(\by_1|\bx_1')\ge
P(\bu_1,\bu_2)W_1(\by_1|\bx_1)
\right\}
\end{eqnarray}
and
\begin{eqnarray}
& &\calA_2(\bu_1,\bu_1',\bu_2,\bu_2',\bx_2,\by_2)\nonumber\\
&=&\calT(Q_2)\bigcap
\left\{\bx_2':~P(\bu_1',\bu_2')W_2(\by_2|\bx_2')\ge
P(\bu_1,\bu_2)W_2(\by_2|\bx_2).
\right\}
\end{eqnarray}
The probability of $\calE(\bu_1,\bu_1',\bu_2,\bu_2',\bx_1,\bx_2,\by_1,\by_2,\calC_1,\calC_2)$
(w.r.t.\ the randomness
of the bin assignment) is given by:
\begin{eqnarray}
& &\overline{\mbox{Pr}}\{\calE(\bu_1,\bu_1',
\bu_2,\bu_2',\bx_1,\bx_2,\by_1,\by_2,\calC_1,\calC_2)\}\nonumber\\
&=&2^{-n(R_1+R_2)}\bigg|[\calC_1\times\calC_2]\bigcap
\calA(\bu_1,\bu_1',\bu_2,\bu_2',\bx_1,\bx_2,\by_1,\by_2)\bigg|+\nonumber\\
&&2^{-n(R_1+R_2)}\bigg|\calC_1\bigcap
\calA_1(\bu_1,\bu_1',\bu_2,\bu_2',\bx_1,\by_1)\bigg|+\nonumber\\
&&2^{-n(R_1+R_2)}\bigg|\calC_2\bigcap
\calA_2(\bu_1,\bu_1',\bu_2,\bu_2',\bx_2,\by_2)\bigg|+\nonumber\\
& &+2^{-n(R_1+R_2)}\calI\{P(\bu_1',\bu_2')\ge P(\bu_1,\bu_2)\},
\end{eqnarray}
where the first term stands for errors of type 5, the second and third terms
represent errors of types 3 and 4, and the last term is associated with an
error where both $\bu_1'\ne\bu_1$ and $\bu_2'\ne\bu_1$, but 
the respective bins both coincide. 
Passing temporarily to shorthand notation, let us denote
\begin{equation}
N\dfn\bigg|[\calC_1\times\calC_2]\bigcap
\calA\bigg|+
\bigg|\calC_1\bigcap\calA_1\bigg|+\bigg|\calC_2\bigcap\calA_2\bigg|+
\calI\{P(\bu_1',\bu_2')\ge
P(\bu_1,\bu_2)\}\dfn N_{12}+N_1+N_2+I.
\end{equation}
The next step, as before, is to average over the randomness of all codewords in $\calC_1$
and $\calC_2$, 
To analyze the large deviations behavior of $N_{12}+N_1+N_2$,
the contributions of the individual random variables can be handled
separately, since 
$\mbox{Pr}\{N_{12}+N_1+N_2> \mbox{threshold}\}$
is of the same exponential order of the sum
$$\mbox{Pr}\{N_{12}> \mbox{threshold}\}+
\mbox{Pr}\{N_{1}> \mbox{threshold}\}+
\mbox{Pr}\{N_{2}> \mbox{threshold}\}.$$
Now, $N_1$ and $N_2$ are binomial random variables
whose numbers of trials are $2^{nR_1}$ and $2^{nR_2}$, respectively, and whose
probabilities of success decay exponentially according to the relevant channel
mutual informations, similarly as before.
So their contributions are again analyzed with great similarly to 
those of type 1 and type 2 errors. 

Finally, it remains to handle $N_{12}$, which not a binomial random
variables, but it can be decomposed as the sum (over combinations of
conditional types of
$\bx_1'$ given $\by_1$ and of $\bx_2'$ given $\by_2$)
of products of independent binomial random variables, for which we reuse the
notations $N_1$ and $N_2$ (for a given combination of such types). 
Using the same techniques as in \cite[Chap.\ 6]{fnt}, one can easily obtain
the following generic result concerning the large deviations behavior of
$N_1\cdot N_2$:
If $N_1$ is a binomial random variable with $2^{nA_1}$ trials and probability of success
$2^{-nB_1}$ and
$N_2$ is an independent binomial random variable with $2^{nA_2}$ trials and probability of success
$2^{-nB_2}$, then
\begin{eqnarray}
\mbox{Pr}\{N_1\cdot N_2 \ge 2^{nC}\}&\exe&\max_{0\le\alpha\le C}
\mbox{Pr}\{N_1\ge 2^{n\alpha}\}\cdot \mbox{Pr}\{N_2\ge
2^{n(C-\alpha)}\}\nonumber\\
&\exe& 2^{-nE}
\end{eqnarray}
with
\begin{equation}
E=\left\{\begin{array}{ll}
[B_1-A_1]_++[B_2-A_2]_+ & C \le [A_1-B_1]_++[A_2-B_2]_+\\
\infty & C > [A_1-B_1]_++[A_2-B_2]_+\end{array}\right.
\end{equation}
Using this fact, it is possible to obtain the contribution of the type 5
error event.

Upon carrying out the analysis along these lines, the state of affairs turns
out to be as described next.
In the analysis of the conditional probability of error,
the contribution of a given type class, $\calT(\bu_1',\bu_2')$, of competing source
vectors, which are encoded
into $\bx_1'$ and $\bx_2'$ (from given conditional type classes given $\by_1$
and $\by_2$, respectively) is the following: the probability of error of type $i$
is of the exponential order of
$\exp\{-n[f_i(\bu_1',\bu_2',\bx_1',\bx_2',\by_1,\by_2)]_+\}$, $i=1,\ldots,5$.
Thus, the total conditional error probability contributed by this combination of types is
of the exponential order of 
\begin{eqnarray}
\sum_{i=1}^5\exp\{-n[f_i(\bu_1',\bu_2',\bx_1',\bx_2',\by_1,\by_2)]_+\}&\exe&
\exp\{-n\min_i[f_i(\bu_1',\bu_2',\bx_1',\bx_2',\by_1,\by_2)]_+\}\nonumber\\
&=&\exp\{-n[f_0(\bu_1',\bu_2',\bx_1',\bx_2',\by_1,\by_2)]_+\}.
\end{eqnarray}
For the total contribution of all type classes, the exponent
$[f_0(\bu_1',\bu_2',\bx_1',\bx_2',\by_1,\by_2)]_+$ should be minimized
over all such combinations of types (that yield the relevant pairwise error event).
An upper bound on this exponent is obtained by selecting the same combination
of types as those of the correct source vectors (instead of taking this
minimum), namely, the conditional error
probability of the 
MAP decoder is simply lower bounded by the exponential order of
$\exp\{-n[f_0(\bu_1,\bu_2,\bx_1,\bx_2,\by_1,\by_2)]_+\}$.
As for the universal decoder, one should minimize the exponent
$[f_0(\bu_1',\bu_2',\bx_1',\bx_2',\by_1,\by_2)]_+$ as well, 
but only over the combinations
of type classes that are associated with the pairwise error event of this
decoder, namely, those for which
$f_0(\bu_1',\bu_2',\bx_1',\bx_2',\by_1,\by_2)\ge
f_0(\bu_1,\bu_2,\bx_1,\bx_2,\by_1,\by_2)$. However, this minimum is exactly
$[f_0(\bu_1,\bu_2,\bx_1,\bx_2,\by_1,\by_2)]_+$, which agrees with that of the upper bound
associated with the MAP decoder. 

More formally, denoting
$f_0(\bu_1,\bu_2,\bx_1,\bx_2,\by_1,\by_2)$ as a functional of the
relevant joint  empirical distributions, i.e.,
\begin{equation}
F(\hat{P}_{\bu_1\bu_2},\hat{P}_{\bx_1\by_1},\hat{P}_{\bx_2,\by_2})\dfn
[f_0(\bu_1,\bu_2,\bx_1,\bx_2,\by_1,\by_2)]_+,
\end{equation}
the error exponent achieved by both the MAP decoder and the universal decoder
is given by
\begin{eqnarray}
E(R_1,R_2,Q_1,Q_2)&=&\min_{P_{U_1'U_2'},W_1',W_2'}\left\{D(P_{U_1'U_2'}\|P_{U_1U_2})+
D(W_1'\|W_1|Q_1)+\right.\nonumber\\
& &\left.D(W_2'\|W_2|Q_2)+
F(P_{U_1'U_2'},Q_1\times W_1',Q_2\times W_2')\right\}.
\end{eqnarray}

Note that when $R_1$ and $R_2$ are sufficiently large, neither
$f_3$ nor
$f_4$ would achieve
$f_0$. At the same time,
$f_1$, $f_2$ and $f_5$
degenerate as follows:
\begin{eqnarray}
f_1(\bu_1,\bu_2,\bx_1,\bx_2,\by_1,\by_2)&=&
\hI(X_1;Y_1)-\hH(U_1|U_2)\\
f_2(\bu_1,\bu_2,\bx_1,\bx_2,\by_1,\by_2)&=&
\hI(X_2;Y_2)-\hH(U_2|U_1)\\
f_5(\bu_1,\bu_2,\bx_1,\bx_2,\by_1,\by_2)&=&
[\hI(X_1;Y_1)+\hI(X_2;Y_2)-\hH(U_1,U_2)]_+.
\end{eqnarray}
Therefore, we have
\begin{eqnarray}
E(\infty,\infty,Q_1,Q_2)&=&\min_{P_{U_1'U_2'},W_1',W_2'}\left[D(P_{U_1'U_2'}\|P_{U_1U_2})+
D(W_1'\|W_1|Q_1)+
D(W_2'\|W_2|Q_2)+\right.\nonumber\\
& &\left.\min\{I(X_1;Y_1')-H(U_1'|U_2'),
I(X_2;Y_2')-H(U_2'|U_1'),\right.\nonumber\\
& &\left. I(X_1;Y_1')+I(X_2;Y_2')-H(U_1',U_2')\}\right].
\end{eqnarray}
Further restricting this to the case of noiseless bit--pipes at fixed
transmission rates $r_1$
and $r_2$, respectively, the above
channel--related divergence terms vanish, and one obtains the error exponent
of separate compression and joint decompression of correlated sources
\begin{equation}
\min_{P_{U_1'U_2'}}\left[D(P_{U_1'U_2'}\|P_{U_1U_2})+
\min\{r_1-H(U_1'|U_2'),
r_2-H(U_2'|U_1'),
r_1+r_2-H(U_1',U_2')\}\right],
\end{equation}
in agreement with \cite[Exercise 13.5]{CK81} (second edition).

\section*{Appendix -- Proof of Eq.\ (\ref{csiszargallager})}
\renewcommand{\theequation}{A.\arabic{equation}}
    \setcounter{equation}{0}

First, observe that
\begin{equation}
(1+\rho)\ln\left(\sum_{u\in\calU}[P(u)]^{1/(1+\rho)}\right)=\max_{P'}[\rho
\calH(P')-D(P'\|P)],
\end{equation}
as can easily be seen by solving explicitly the maximization on the r.h.s.
Next observe that for $\rho> 0$, the maximizer $P_0'$ is always associated with an entropy
larger\footnote{To see why this is true, note that $\rho\calH(P)\le\max_{P'}[
\rho\calH(P')-D(P'\|P)]=\rho\calH(P_0')-D(P_0'\|P)\le\rho\calH(P_0')$.}
than $\calH(P)=H(U)$. 
Therefore, the above identity can be further developed to obtain
\begin{eqnarray}
& &(1+\rho)\ln\left(\sum_{u\in\calU}[P(u)]^{1/(1+\rho)}\right)\nonumber\\
&=&\max_{\{P':~\calH(P)\le\calH(P')\le\log|\calU|\}}[\rho\calH(P')-D(P'\|P)]\nonumber\\
&=&\max_{\calH(P)\le \tR\le\log|\calU|}\max_{\{P':~\calH(P')=\tR\}}[\rho\calH(P')-D(P'\|P)]\nonumber\\
&=&\max_{\calH(P)\le \tR\le\log|\calU|}\max_{\{P':~\calH(P')=\tR\}}[\rho \tR-D(P'\|P)]\nonumber\\
&=&\max_{\calH(P)\le \tR\le\log|\calU|}\left[\rho
\tR-\min_{\{P':~\calH(P')=\tR\}}D(P'\|P)\right]\nonumber\\
&=&\max_{\calH(P)\le \tR\le\log|\calU|}[\rho
\tR-E^{\mbox{\tiny s}}(\tR)],
\end{eqnarray}
where the last step follows from the fact that, due to the convexity and the
monotonicity of the source coding exponent function, the constraint
$\calH(P')\ge \tR$, of the
minimization of $D(P'\|P)$ that defines it, is attained with equality
in the range $\tR\in[\calH(P),\log|\calU|]$ (see also \cite[eq.\ (7)]{Csiszar80}).
Therefore,
\begin{eqnarray}
& &\max_{0\le\rho\le 1}\left\{E_0(\rho,Q)-
(1+\rho)\ln\left(\sum_{u\in\calU}[P(u)]^{1/(1+\rho)}\right)\right\}\nonumber\\
&=&\max_{0\le\rho\le 1}\left[E_0(\rho,Q)-\max_{\calH(P)\le
\tR\le\log|\calU|}[\rho \tR-E^{\mbox{\tiny s}}(\tR)]\right\}\nonumber\\
&=&\max_{0\le\rho\le 1}\min_{\calH(P)\le \tR\le\log|\calU|}[E_0(\rho,Q)-\rho
\tR+E^{\mbox{\tiny s}}(\tR)]\nonumber\\
&=&\min_{\calH(P)\le \tR\le\log|\calU|}\max_{0\le\rho\le 1}[E_0(\rho,Q)-\rho
\tR+E^{\mbox{\tiny s}}(\tR)]\nonumber\\
&=&\min_{\calH(P)\le \tR\le\log|\calU|}[E_{\mbox{\tiny r}}^{\mbox{\tiny c}}(\tR,Q)
+E^{\mbox{\tiny s}}(\tR)],
\end{eqnarray}
where the interchange of the maximization and the minimization in the second
to the last step is allowed by the concavity of $E_0(\rho,Q)$ in $\rho$
\cite[eq.\ (5.6.26)]{Gallager68} and the convexity of $E^{\mbox{\tiny s}}(\tR)$
\cite{Csiszar80}.

\section*{Acknowledgement} Interesting discussions with Jacob Ziv are
acknowledged with thanks.

\end{document}